\documentclass[reprint,superscriptaddress,showpacs,showkeys,amsmath,amssymb,aps,nofootinbib,preprintnumbers]{revtex4-1}[12pt]

\usepackage{graphicx}
\usepackage[mathlines]{lineno}
\usepackage[english]{babel}
\usepackage{color}
\usepackage{ulem}
\usepackage{tikz}

\def\lapproxeq{\lower .7ex\hbox{$\;\stackrel{\textstyle
<}{\sim}\;$}}
\def\gapproxeq{\lower .7ex\hbox{$\;\stackrel{\textstyle
>}{\sim}\;$}}

\makeatletter
\def\myline{\pgfutil@ifnextchar[{\my@line}{\my@line[]}}%
\def\my@line[#1](#2)(#3){%
\tikz[remember picture,overlay] \draw[#1]  (#2)--(#3); 
}%
\begin{document}

\preprint{Submitted to Phys. Rev. D} 

\date{\today}

\title{Askaryan radiation from neutrino-induced showers in ice}
\author{Jaime Alvarez-Mu\~niz}
\affiliation{%
Instituto Galego de F\'\i sica de Altas Enerx\'\i as (IGFAE),
Universidade de Santiago de Compostela, 15782 Santiago
de Compostela, Spain
}%
\author{P. M. Hansen}
\affiliation{%
Instituto de  F\'isica de La Plata (CCT La Plata-CONICET) \& 
Departamento de F\'isica, Facultad de Ciencias Exactas  \&  
Departamento de Ciencias B\'asicas, Facultad de Ingenier\'ia,
Universidad Nacional de La Plata,
C. C. 67 - 1900 La Plata, Argentina
}%
\author{Andr\'es Romero-Wolf}
\affiliation{%
Jet Propulsion Laboratory, California Institute of Technology, 4800 Oak Grove Drive,
Pasadena, California 91109, USA
}%
\author{Enrique Zas}
\affiliation{%
Instituto Galego de F\'\i sica de Altas Enerx\'\i as (IGFAE),
Universidade de Santiago de Compostela, 15782 Santiago
de Compostela, Spain
}%
\date{\today}

\begin{abstract}
We present a semi-analytical method for the calculation of coherent Askaryan radiation in showers induced by neutrinos of any flavor in ice. 
We compare our results with those of a full Monte Carlo simulation based on the ZHAireS code. This approach is able to reproduce the vector potential and hence electric field at any experimentally relevant observer position in the time domain. This work extends published results only valid for electron-induced showers. We establish the validity of the semi-analytical calculation of the radio signal produced by all types of neutrino-induced showers in ice. The method is computationally efficient and only requires as inputs the longitudinal charge excess profile of the showers and a parameterization of the vector potential in the far-field region of the shower at the Cherenkov angle 
that we also provide. Our methodology avoids tracking the contributions to the electric field from millions of particles every time the radio pulse has to be calculated at a given observer position. These results can be readily used in the interpretation of the data taken by experiments, and in the planning and design of future initiatives based on the radio technique in ice. 
\end{abstract}



\maketitle

\def\linenumberfont{\normalfont\tiny\itshape\sffamily}

\section{Introduction}

Ultra-high energy (UHE) cosmic neutrinos at EeV energies and above are quite possibly the unique messengers of the most distant, energetic and powerful sources in the universe. Neutrinos travel through space undeflected in the galactic and extra-galactic magnetic fields and can traverse unaffected through regions of space with sizable matter depths. Their detection would allow us to probe regions of the universe hidden to conventional photon astronomy. The search for UHE neutrino fluxes from transient objects observed in gravitational waves has already proved to be extremely rewarding, contributing to the leap forward that resulted from the correlation between a neutron star merger~\cite{NS_Merger2017} and the subsequent {\it kilonova} that was also detected throughout the electromagnetic spectrum~\cite{NSFollowUp}, and followed up in neutrinos with no candidates found~\cite{NS_NuSearch}. Moreover, UHE-neutrino detection has the potential to provide answers to long-standing questions on the origin, nature and production mechanisms of the UHE cosmic rays \cite{Nagano-Watson_2000,Dawson_2017,Mollerach-Roulet_2018} and its relation with the astrophysical neutrino flux already detected by Icecube \cite{IceCube_PRL2014,IceCube_ApJ_2016}. 

The detection of UHE neutrinos is experimentally challenging due to the low fluxes expected and to the small probability of neutrino interaction with matter \cite{Becker_PhysRep_2008,Anchordoqui_review_2014}. A variety of techniques are being exploited for this purpose, see \cite{Veronique_review_2011, Alvarez-Muniz_ICRC17} for reviews, namely, arrays of photo-multiplier tubes buried in ice or under water that observe the Cherenkov light produced by showers or particle tracks induced by neutrinos, particle arrays that sample the front of atmospheric showers induced by $\nu$, arrays of antennas measuring radiopulses from air showers~\cite{AERA,GRAND} and, most relevant to this paper, arrays of antennas in dense, dielectric media \cite{ARIANNA_2014,ARA_PRD2016,Connolly-Vieregg_Review_2017}. The latter are designed to detect the radiation in the MHz-GHz frequency range generated in $\nu$-induced showers from the excess charge that develops in showers, first discussed by Askaryan in 1962~\cite{Askaryan62}. Despite all of these efforts neutrinos have escaped detection in the EeV energy range by existing experiments using these techniques \cite{Auger_nus_JCAP2019,Auger_point_JCAP2019,IceCube_PRD2018,ANITA_2019}. About three orders of magnitude in energy below, in the 100 TeV to PeV range, the IceCube experiment has detected with high confidence a flux of astrophysical neutrinos \cite{IceCube_PRL2014,IceCube_ApJ_2016}, including evidence of an energetic neutrino from a direction consistent with blazar TXS 0506+056 in temporal coincidence with a flaring state \cite{IceCube_TXS}.

Radio emission of particle showers induced by a cosmic ray or a neutrino in a dense, dielectric media is mainly due to the electromagnetic component which develops an excess negative charge producing  coherent radio emission at wavelengths longer than the size of the emitting region \cite{Askaryan62,ZHS92}. The electric field in the MHz to GHz frequency range increases linearly with frequency up to a characteristic cut-off, of few GHz in a dense medium such as ice, and the emitted power in radiowaves scales with the square of the particle energy \cite{Askaryan62,ZHS92}. These predictions have been confirmed in accelerator experiments \cite{Saltzberg_SLAC_sand,Gorham_SLAC_salt,Gorham_SLAC_ice}.
Askaryan radiation is known to be directly related to the time-variation of the net charge of the shower~\cite{ARZ10, ARZ11}, inducing a complicated bipolar electric pulse with a frequency spectrum which is angular dependent. It lasts a few nanosecond in the Cherenkov direction in a dense medium, and has been shown to be in good agreement with experimental observations \cite{Miocinovic_SLAC_sand}. 
These findings have motivated a variety of experiments to search for these pulses using Antarctic ice \cite{RICE03,ANITA_long_2009,ARIANNA_2014,ARA_PRD2016,Connolly-Vieregg_Review_2017} and the surface of the Moon~\cite{LUNASKA1,GLUE,Zhelez_moon,LUNASKA2,RESUN,Westerbork,Parkes,Bray_APP_2016} as targets.

The success of these initiatives and the exploration of new ones, requires an accurate and computationally efficient calculation of the properties of Askaryan radiation in UHE showers. Several approaches have been pursued. Detailed Monte Carlo (MC) simulations of UHE showers and the associated radio emission in dense media have been developed \cite{ZHS92,ZHAireS_ice}. With this approach, the full complexity of shower phenomena is accounted for, including the inherent shower-to-shower fluctuations and the elongation of the showers at UHE due to the Landau-Pomeranchuk-Migdal effect \cite{LPM,Stanev_LPM}. However, MC methods are typically very time-consuming at UHE. For instance, a ZHAireS \cite{ZHAireS_ice} simulation of the radiation produced in a $E=1$ EeV neutrino-induced shower at six different observer positions takes on the order of 2 hours 
of CPU time using a thinning energy $\sim 10^{-5}E$, with the CPU time scaling linearly with the number of observers. 
Analytical techniques have also been applied \cite{Buniy_PRD2001,Hanson_APP2017}, in which parameterizations of the longitudinal and lateral profile of the shower are used to model the space-time evolution of the excess charge distribution, and used as input to Maxwell's equations. These methods are computationally efficient, but ignore the large shower-to-shower fluctuations due to the LPM effect~\cite{alz97,alz98,alvz99} which cannot be easily parameterized~\cite{Hanson_APP2017} and result in qualitative differences in the electric field impulse for energies above $\sim 10^{17}$ eV in ice~\cite{ARZ10}. Some efforts have been also made to calculate the pulses solving Maxwell's equations directly with finite difference time-domain methods~\cite{Chen_2012}. Although these methods are also capable of producing very accurate results they are computationally intensive, typically much more than MC simulations. 

Semi-analytical methods have been shown to be a very good compromise between MC and analytical techniques \cite{ARZ11}. The idea in this case is to obtain an approximate charge distribution from detailed MC simulations to be used as the input for analytical calculations. In \cite{ARZ11} a semi-analytical calculation was presented and demonstrated to reproduce the electric field in the time domain at all angles with respect to shower axis in both the Fraunhofer and Fresnel zones. The shower-to-shower fluctuations and the influence of the LPM effect~\cite{alz97,alz98,alvz99} can be accurately accounted for with the MC simulation of the shower profile, what makes this technique accurate and computationally efficient. However, in \cite{ARZ11} such a semi-analytical approach was developed only for purely electromagnetic showers. In this work the model is extended to hadronic showers, and shown to be accurate for the calculation of the time-domain electric field produced in UHE showers induced by neutrinos of any flavor in both charged-current (CC) and neutral-current (NC) interactions, including those showers produced by the decay products of secondary $\tau$ leptons produced in charged-current $\nu_\tau$ interactions.

This paper is organized as follows. In Section \ref{S:Semi} we present a summary of the semi-analytical method given in more detail in \cite{ARZ11}. In Section \ref{S:TheSimulations} we report on the simulations of electron, proton and neutrino-induced showers used in this work. We give our results, and comparisons of our semi-analytical approach with full MC simulations in Section \ref{S:Results}. In Section \ref{S:Conclusions} we summarize and conclude the paper.
 
\section{Askaryan radiation}
\label{S:Semi}

In this Section, we review the calculation of the electric field (Askaryan radiation) due to the charge excess of a generic shower, characterized by the longitudinal and lateral shower development (respectively parallel and perpendicular to the shower axis), in a dielectric medium such as ice. For completeness and self-consistency of the paper we summarize here the procedure explained in \cite{ARZ11}, where full details can be found. 

\subsection{General formalism}

\begin{figure}[ht]
{\centering 
\resizebox*{0.51\textwidth}{!}{\includegraphics{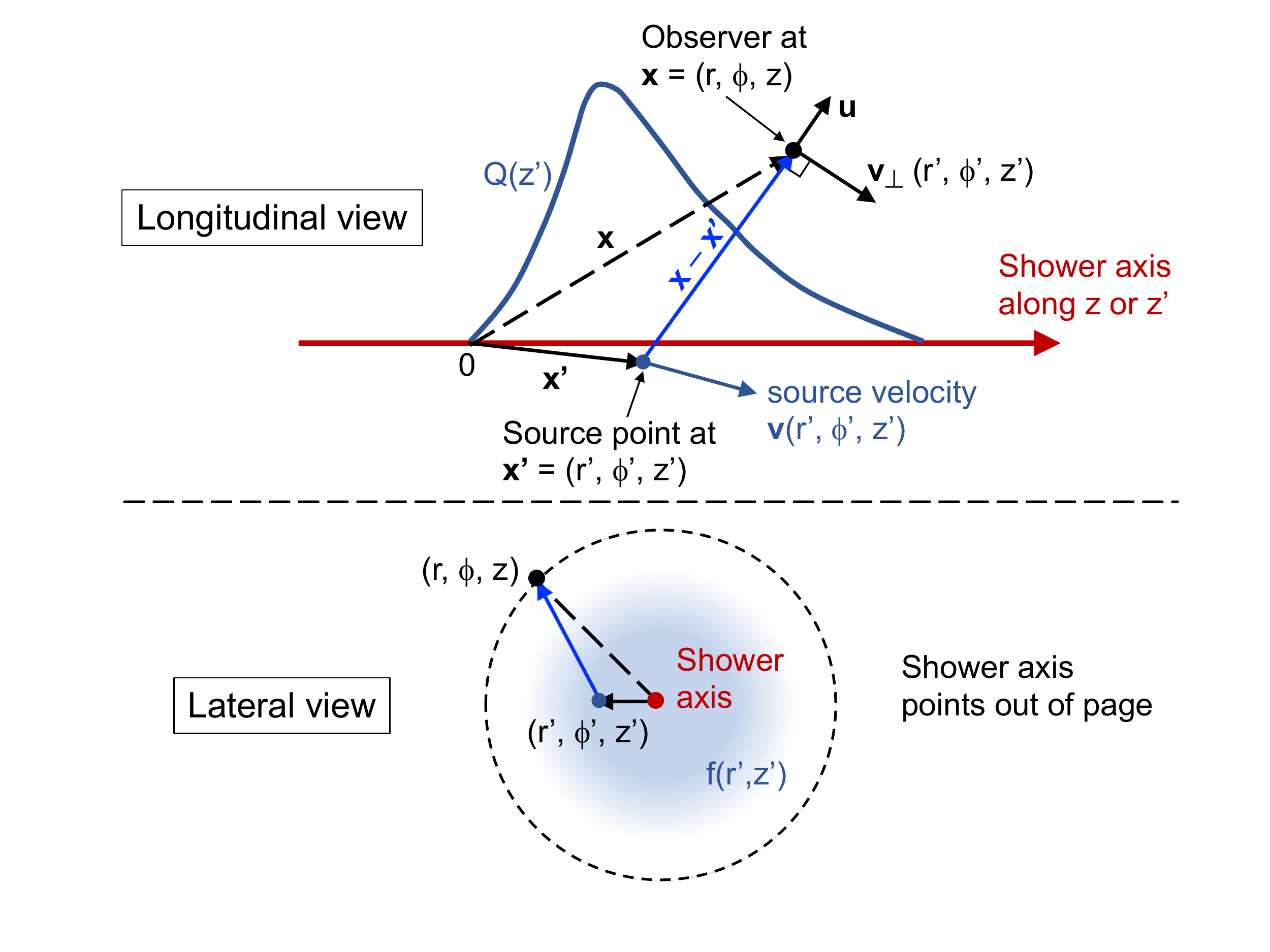}}
\par}
\caption{Sketch of the geometry of a particle shower. In the top panel we show the longitudinal (side) 
view along the shower axis that is parallel to $z$ (or $z'$). In the bottom panel we show the lateral (frontal) 
view in the plane perpendicular to the shower axis. The observer and source positions are indicated 
with the vectors ${\mathbf x}=(r,\phi,z)$ and ${\mathbf x}'=(r',\phi, z)$, both in cylindrical coordinates. 
The source travels at a velocity ${\mathbf v}(r',\phi',z')$ with ${\mathbf v}_\perp$ the component of 
${\mathbf v}$ in the direction perpendicular to the unit vector ${\mathbf{\hat u}}$ along 
the observer direction (${\mathbf x} - {\mathbf x}'$). The longitudinal development of the charge excess 
is denoted as $Q(z')$, while the lateral spread is denoted as $f(r',z')$. }
\label{fig:geometry}
\end{figure}

Let us consider a charge distribution $\rho({\mathbf x}')$ where ${\mathbf x}'$ denotes the source position, traveling at velocity $\mathbf{v}$, with current density vector $\mathbf{J}=\rho\mathbf{v}$.
The Green's function solutions to Maxwell's equations provide the potentials $\Phi({\mathbf x},t)$ and $\mathbf{A}({\mathbf x},t)$ with ${\mathbf x}$ and $t$ the observer position and time respectively (see Fig.~\ref{fig:geometry} for a sketch of the geometry).  In the Coulomb gauge ($\nabla\cdot\mathbf{A}=0$) the solutions can be written as \cite{Jackson}:  
\begin{equation}
\Phi(\mathbf{x},t)=\frac{1}{4\pi\epsilon}\int_{-\infty}^{\infty}  
\frac{\rho(\mathbf{x}',t)}{|\mathbf{x}-\mathbf{x}'|} \,d^3\mathbf{x}' ,
\end{equation} 
\begin{equation}
\begin{split}
&
\mathbf{A}(\mathbf{x},t)=\frac{\mu}{4\pi}\int_{-\infty}^{\infty}  
\frac{\mathbf{J}_{\perp}(\mathbf{x}',t')}{|\mathbf{x}-\mathbf{x}'|}
\\
&
\delta\left(\sqrt{\mu\epsilon}\,|\mathbf{x}-\mathbf{x}'|-(t-t')\right) d^3\mathbf{x}' dt' ,
\end{split}
\label{VectorPotential}
\end{equation} 
where $\epsilon$ and $\mu$ are the dielectric and magnetic constants of the medium. The Dirac $\delta-$function relates the observer's time $t$ and the source time $t'$ through the time it takes light to reach the observation point $\mathbf{x}$ from the source position at $\mathbf{x}'$. In the Coulomb gauge, only the transverse component of the current density is relevant \cite{Jackson_AmJPhys}, and it is given by  $\mathbf{J}_{\perp}=-\mathbf{\hat{u}}\times\-(\mathbf{\hat{u}}\times\mathbf{J})$ where $\mathbf{\hat{u}}=(\mathbf{x}-\mathbf{x}')/|\mathbf{x}-\mathbf{x}'|$ is a unit vector pointing from the source position $\mathbf{x}'$ to the observer at $\mathbf{x}$. In the Coulomb gauge, the scalar potential only describes reactive near-field terms which will be ignored for our purposes. This simplifies the computation of the radiative electric field $\mathbf{E}=-\nabla\Phi-\partial\mathbf{A}/\partial t$ to a time derivative 
$\mathbf{E}=-\partial\mathbf{A}/\partial t$. This has been shown to be a valid approximation for distances to shower axis in excess of a meter and frequencies above 10 MHz~\cite{ZHS_calculations_PRD2013}. 

We first approximate the excess charge distribution as a flat pancake traveling with velocity ${\mathbf v}$ along the shower axis parallel to $z'$. 
The associated current density can be assumed to have cylindrical symmetry and in that case it can be generically written in cylindrical coordinates as \cite{ARZ11}:
\begin{equation}
\mathbf{J}(\mathbf{x}',t')=\mathbf{v}(r',\phi',z')f(r',z')Q(z')\delta(z'-vt'),
\label{eq:current}
\end{equation}
where $r'=\sqrt{x'^2+y'^2}$ is the radial distance and $\phi'$ the azimuth angle. $Q(z')$ gives the distribution of the excess charge along the shower axis, parallel to $z'$. The function $f(r',z')$ represents the lateral charge distribution in a plane transverse to the $z'$ axis, that depends on the stage of shower development along $z'$. The velocity $\mathbf{v}(r',\phi',z')$ is mainly parallel to $z'$ but it can have a radial component due to scattering of particles in the shower and transverse momenta acquired in interactions. 

With the approximations for the current density given in Eq.\,(\ref{eq:current}) the vector potential in Eq.\,(\ref{VectorPotential}) becomes:
\begin{equation}
\begin{split}
&
\mathbf{A}(\mathbf{x},t)=\frac{\mu}{4\pi}
\int^{\infty}_{-\infty} dt'
\int^{\infty}_{-\infty} dz' Q(z') \,\delta(z'-vt')
\\
& 
\int^{\infty}_{0} dr' r' \int^{2\pi}_{0} d\phi'  f(r',z') \,\mathbf{{v}}_{\perp}(r',\phi',z') 
\\
&
\frac{\delta\left(n|\mathbf{x}-\mathbf{x}'|/c-(t-t')\right)}{|\mathbf{x}-\mathbf{x}'|} ,
\end{split}
\label{eq:vp}
\end{equation}
where we have used that $c/n=1/\sqrt{\mu\epsilon}$, with $n$ the refractive index and $c$ the speed of light. Here $\mathbf{v}_{\perp}(r',\phi',z')$ is the velocity projected along the plane perpendicular to the observer direction ${\mathbf{\hat u}}$. Eq.\,(\ref{eq:vp}) can be cast in cylindrical coordinates, the observer being at ${\mathbf x}=(r\cos\phi,r\sin\phi,z)$ and with an equivalent expression for ${\mathbf x'}$. The distance  $|\mathbf{x}-\mathbf{x}'|$ is then:
\begin{equation}
|\mathbf{x}-\mathbf{x}'|= \sqrt{r^2+r'^2-2rr'\cos\phi'+(z-z')^2} ,
\label{eq:Fresnel_approx}
\end{equation}
where, given the cylindrical symmetry of the model for the current density, and for simplicity, we have assumed (without loss of generality) that the observer is placed at $\phi=0$.

We note that the approximation in Eq.~\ref{eq:current} neglects the curvature of the shower front. The validity of this and other approximations (see the following subsection) will be ultimately justified in Section~\ref{S:Results}, where we present a comprehensive comparison between the vector potential obtained in full simulations that account for all the details of the evolution of the lateral density with shower depth, and that predicted by the model discussed in this article. We can think of the whole approach as a means to obtain a useful parameterization of the radio emission which is totally justified in view of its results. 

\subsection{Factorization of the lateral distribution}

We now consider the case of an observer placed in a region in which the Fraunhofer approximation is valid for the source contributions at a fixed $z'$ position. This requires that the observation distance satisfies $R >> r'^2/\lambda$, where $r'$, the lateral dimension of the shower, is typically $< 1$ m in ice, implying $R >> 10$ m for frequencies up to 3~GHz. Note that the Fraunhofer approximation is not necessarily valid for the whole emission region, since that would require a similar relation involving the longitudinal dimension of the shower instead of $r'$, which will restrict the validity to much greater distances~\cite{ZHS_calculations_PRD2013}. Since the typical separation in planned and under development antenna arrays in ice is at least 10 m~\cite{Connolly-Vieregg_Review_2017}, we expect this approximation to be valid for most practical situations. 
Under the Fraunhofer approximation, we expand Eq.~(\ref{eq:Fresnel_approx}) to first order in $r'$ and obtain: 
\begin{equation}
|\mathbf{x}-\mathbf{x}'|= \sqrt{r^2+(z-z')^2}-r'\sin\theta(z')\cos\phi' ,
\label{eq:R_Fresnel}
\end{equation}
where $\theta(z')$ is the {\it local} observation angle, that is the angle between the vector $\mathbf{\hat{u}}_0(z')$, pointing from $\mathbf{x}'=(r'=0,\phi'=0,z')$ (with $z'$ at the shower axis) to the observer, given by 
$\sin\theta(z') = r/\sqrt{r^2+(z-z')^2}$. 

If we also make the assumption that the shape of the lateral density $f(r',z')$ 
depends weakly on $z'$, we can drop the $z'$ dependence in Eq.\,(\ref{eq:current}), 
and then Eq.\,(\ref{eq:vp}) can be integrated in $t'$ to give the vector potential as: 
\begin{equation}
\begin{split}
&
\mathbf{A}(r,z,t)= \frac{\mu}{4\pi}
\int_{-\infty}^{\infty} dz'\frac{Q(z')}{\sqrt{r^2+(z-z')^2}} 
\\
&
\int_{0}^{\infty} dr' r' \int_{0}^{2\pi} d\phi' f(r') ~\frac{\mathbf{v}_{\perp}(r',\phi',z')} {v} 
\\
& 
\delta\left(\frac{z'}{v} + \frac{n\sqrt{r^2+(z-z')^2}-nr'\sin\theta(z') \cos\phi'}{c}-t\right).
\label{eq:vp_Fresnel}
\end{split}
\end{equation}
The resulting vector potential at position $(r,z)$ and observer's time $t$, is a sum of the vector potentials produced by the current density
${\mathbf J}(r',z',t')$, where the space-time position of the charge $(r',\phi',z',t')$ is constrained by the argument of the $\delta$-function in Eq.\,(\ref{eq:vp_Fresnel}), accounting for the well known retarded time $t'$, such that $t-t'$ is the light travel time from source to observer. 
We note that the resulting space integral in $r'$ and $\phi'$ given by the last two lines in Eq.~(\ref{eq:vp_Fresnel}) depends only on $z'$ and it can be factored out in the expression. The simplicity of the model is directly related to this fact.  

Following the same procedure as in \cite{ARZ11}, we now define a {\it form factor} ${\mathbf F}$, corresponding to the $r',\phi'$ integrals in Eq.\,(\ref{eq:vp_Fresnel}) as: 
\begin{equation}
\begin{split}
&
{\mathbf F}\left(t-\frac{z'}{v}-\frac{n\sqrt{r^2+(z-z')^2}}{c}\right) =
\\
& 
\int_{0}^{\infty} dr'r' \int_{0}^{2\pi} d\phi' ~ f(r') ~\frac{\mathbf{v}_{\perp}(r',\phi',z')}{v}
\\
& 
\delta\left(\frac{z'}{v} + \frac{n\sqrt{r^2+(z-z')^2}}{c}-t-\frac{nr'\sin\theta(z')\cos\phi'}{c}\right) .
\end{split}
\label{eq:lambda_Fresnel}
\end{equation} 
The form factor ${\mathbf F}$ is a function that depends on the shower structure in the medium and accounts for the interference effects due to the lateral spread \cite{Buniy_PRD2001}. Using the same procedure as in \cite{ARZ11}, ${\mathbf F}$ can be decomposed in a component along the direction $\hat{\mathbf u}_0(z')$ (from shower axis to the observer) which is expected to be rather small because the integral involves $\mathbf{v}_{\perp}$, and a component orthogonal to it, with unit vector $\hat{\mathbf p}_0(z')$:
\begin{equation}
\mathbf{F}(z')=F_p(z') ~\mathbf{\hat{p}}_0(z') + F_u(z') \mathbf{\hat{u}}_0(z')\,.
\label{eq:Fcomponents}
\end{equation} 
Askaryan radiation is mainly polarized in the direction 
$\hat{\mathbf p}_0(z')$ and the component along $\hat{\mathbf u}_0(z')$ can be neglected as shown in \cite{ARZ11}, and confirmed in Section \ref{S:Results} with full Monte Carlo simulations of the electric field that take that component into account.

With the definition of the form factor in Eq.\,(\ref{eq:lambda_Fresnel}), and neglecting the $F_u$ component in Eq.\,(\ref{eq:Fcomponents}), the vector potential reads:
\begin{equation}
\begin{split}
&
\mathbf{A}(r,z,t)=  
\frac{\mu}{4\pi}
\int_{-\infty}^{\infty}
dz'\frac{Q(z')}{\sqrt{r^2+(z-z')^2}}~\mathbf{p}_0(z') 
\\
&
F_p\left(t-\frac{z'}{v}-\frac{n\sqrt{r^2+(z-z')^2}}{c}\right) ,
\label{eq:model_Fres}
\end{split}
\end{equation}
where it becomes apparent that the vector potential in the radiative region of the shower can be obtained as a convolution of the form factor $F_p$ - that effectively accounts for the interference effects due to the lateral distribution of the shower - and the longitudinal profile of the excess charge \cite{ARZ11}.

\subsection{The form factor}

Potentially, the form factor $\mathbf{F}$ could be obtained analytically, although this would be a daunting task. 
Here we follow the same approach as in \cite{ARZ11} which consists on extracting an approximation for $\mathbf{F}$ from simulations of the vector potential $A(\theta_C,t)$ in the Fraunhofer limit and in the Cherenkov direction. It can be  shown~\cite{ARZ11} that the delta function in the integrand of Eq.~(\ref{eq:lambda_Fresnel}) for an observer at a large distance $R$ and in the Fraunhofer limit, can be rewritten so that the expression for the form factor can be casted as: 
\begin{equation}
\begin{split}
&
{\mathbf F}\left(t-\frac{nR}{c}-z'\left[\frac{1}{v}-\frac{n\cos \theta(z')}{c}\right]\right) =
\\
& 
\int_{-\infty}^{\infty} dr'r' \int_{0}^{2\pi} d\phi' ~ f(r') ~\frac{\mathbf{v}_{\perp}(r',\phi',z')}{v}
\\
& 
\delta\left(\frac{nR}{c}-t-\frac{nr'\sin\theta(z')\cos\phi'(z')}{c}+
z'\left[\frac{1}{v}-\frac{n\cos \theta(z')}{c}\right]\right).
\end{split}
\label{eq:lambda_Fraunhofer}
\end{equation} 
The explicit dependence of $\theta$ and $\phi$ on $z'$ can be disregarded in the Fraunhofer limit. 

Evaluating the vector potential in the Cherenkov direction, so that the factor multiplying $z'$ in the delta function in Eq.~(\ref{eq:lambda_Fraunhofer}) vanishes, the form factor integral factorizes from the $z'$ integral of Eq.~\ref{eq:vp_Fresnel} which can then be simply expressed as: 
\begin{equation}
\mathbf{A}(r,z,t)= \frac{\mu}{4\pi R} 
{\mathbf F}\left(t-\frac{nR}{c}\right)
\int_{-\infty}^{\infty} dz'Q(z') .
\label{eq:F-Factorization}
\end{equation}
The form factor is directly proportional to the vector potential and the proportionality factor is the integral of the charge excess over $z'$, which is often referred as the {\it excess projected track-length} \cite{ZHS92}, which we denote as $LQ_{\rm tot}$. This can be intuitively understood because
an observer located at the Cherenkov angle in the Fraunhofer limit ``sees'' the whole longitudinal shower development along $z'$ at once. The width of the pulse is related to the lateral distribution, what corresponds to destructive interference setting in at a few GHz frequency in ice~\cite{ZHS92}. 

It is a trivial matter to numerically obtain the functional form of the form factor. We obtain the vector potential in the Fraunhofer limit and in the Cherenkov direction of a simulated shower in ice, using well-tested MC codes \cite{ZHAireS_ice} - see also Section~\ref{S:TheSimulations}, and use it to obtain the form factor directly from Eq.~(\ref{eq:F-Factorization}). After projecting on the $\mathbf{\hat{p}}$ direction and neglecting the contribution proportional to $\mathbf{\hat{u}}$~\footnote{$\mathbf{\hat{p}}_0$ and $\mathbf{\hat{p}}$ coincide in the Fraunhofer limit and the same happens with $\mathbf{\hat{u}}_o$ and $\mathbf{\hat{u}}$.} the dominant component of the form factor becomes:
\begin{equation}
F_p\left(t-\frac{nR}{c}\right) = \frac{4\pi}{\mu}~\frac{RA(\theta_C,t)}{LQ_{\rm tot}}~
\frac{1}{\sin\theta_C} .
\label{eq:lambda}
\end{equation}

In Sections \ref{S:TheSimulations} and \ref{S:Results}, we describe the simulations performed to extract the form factor. We are interested in showers initiated by neutrinos of all flavors through charged-current and neutral-current interactions, including those showers produced by the secondary products of $\tau$-lepton decays induced in $\nu_\tau$ CC interactions. 

We will show that it is sufficient to calculate two different types of form factors for an accurate calculation of the Askaryan radiation in all these showers, which can be used in Eq.~(\ref{eq:model_Fres}) to accurately obtain the vector potential in the radiative zone, at relatively small distances to the shower, down to order 10~m. The only ingredient that is needed is the longitudinal profile of the excess charge. This methodology allows swift calculations of the radio pulse in most regions of interest for experimental facilities.

This method works because the lateral distribution has been assumed to be independent of $z'$, which is just an approximation. In view of the good description, obtained and discussed later, the developed procedure gives an effective form factor that must account for some convenient averaging of the lateral distribution. 

\section{Neutrino-induced showers in ice}
\label{S:TheSimulations}

\subsection{Neutrino interactions}
\label{S:First}

We consider the following neutrino-nucleon ($\nu N$) interactions at EeV energies: 

\begin{center}
\renewcommand{\arraystretch}{1.3}
\begin{tabular}{ l l }
$\nu_e+N \rightarrow e^{-}+{\rm jet}$  & \hspace{-3cm} $\nu_e$ Charged-Current~(CC) \\
\\
$\nu_X+N \rightarrow \nu_{X}+{\rm jet}$  &\hspace{-3cm} $\nu_X$ Neutral-Current~(NC) \\
where $X\,=\,e,\,\mu\,~{\rm or}~\tau$ & \\ 
\\
$\nu_\tau+N \rightarrow \tau^{-}+{\rm jet}$  &\hspace{-3cm} $\nu_\tau$ Charged-Current~(CC) \\
\\
where $\tau^-\rightarrow
\begin{cases}
~e^- + \nu_\tau + \bar\nu_e~~~\sim\,17\% \\
~\mu^- + \nu_\tau + \bar\nu_\mu~~~\sim\,17\% \\
~{\rm hadrons} + \nu_\tau~~~\sim\,56\% \\
\end{cases}$ & \\
\end{tabular}
\end{center}
and the corresponding interactions for anti-neutrinos which are essentially indistinguishable at UHE \cite{Connolly_nu_xsection_PRD2011}.

Here $N$ represents a nucleon ($p$ or $n$) and the ``jet" represents the secondaries produced in the fragmentation of the nucleon. The $\tau$ lepton produced in a $\nu_\tau$ CC interaction can decay into electrons, muons or hadrons (mainly charged and neutral pions and kaons), with the approximate branching ratios indicated above. 

The $\nu_e$ CC, $\nu_X$ NC and $\nu_\tau$ CC interactions were simulated with HERWIG \cite{HERWIG} to obtain the secondary particles along with their energies and momenta. The decays of the $\tau$ lepton were simulated with TAUOLA \cite{TAUOLA}, giving also decay particles, their energies and their momenta. The secondary particles were then injected (in each case) in the ZHAireS code~\cite{ZHAireS_ice} (see Section~\ref{S:ZHAireS}), a detailed MC program that simulates the subsequent showers and calculates the associated radio emission in homogeneous ice. 

\subsection{Shower simulations}
\label{S:ZHAireS}

Showers simulations in ice were performed with the ZHAireS code \cite{ZHAireS_ice}. ZHAireS is based on the well-known AIRES code \cite{AIRESManual} which we have used in combination with the TIERRAS \cite{TIERRAS} package to simulate showers in dense media, such as ice. Algorithms to calculate the Fourier components of the electric field  produced by charged particle tracks in the shower as well as the electric field in the time domain were implemented in ZHAireS \cite{ZHAireS_ice, ZHAireS_air}. 
These are the same algorithms as first used in the well-known and well-tested Monte Carlo simulations of electromagnetic showers in ice by Zas, Halzen and Stanev with the so-called ZHS code~\cite{ZHS92,ARZ10}. All simulations were performed in homogeneous ice (density $\rho$ = 0.924 g cm$^{-3}$ and refractive index n=1.78, Cherenkov angle $\theta_{c} \sim 55.8^{\circ}$). The thinning level used in the simulations is $10^{-5}$. The results obtained in this paper are restricted to ice, but can be easily extended to other dielectric and homogeneous media such as sand or salt \cite{Saftoiu_salt_2019}. 

By inspecting the interactions enumerated in Section \ref{S:First}, there are basically three different types of showers that are produced in a $\nu N$ interaction at ultra-high energy:
\begin{enumerate}
    \item {\it Mixed} showers: produced in the CC interaction of $\nu_e$. These are composed of a shower initiated by an electron carrying an energy $(1-y)E_\nu$, with $E_\nu$ the primary neutrino energy and $y$, the fraction of energy transferred to the nucleus, and a hadronic shower produced by the fragmentation products of the struck nucleon that carry an energy $yE_\nu$. The two showers are produced almost simultaneously, and their axes are almost parallel at ultra-high energy.
    \item Purely hadronic showers: produced by the secondary products of the struck nucleon in $\nu_X N$ NC interactions as well as in $\nu_\mu$ and $\nu_\tau$ CC interactions, neglecting the possible secondary interactions of the high-energy muons either produced directly in the interaction or in the decay of the $\tau$ lepton. 
    \item {\it Double} showers: produced in the CC interaction of $\nu_\tau$ and the subsequent decay of the $\tau$ lepton. In this case the showers can interfere or not depending on how far apart they are. Given that the $\tau$-decay length is $L_\tau \sim 5\,{\rm km}\,(E_\tau/10^{17}\,{\rm eV})$, and the showers are just tens of meters in length, they can be considered to be independent in most cases. For lower energy showers it would be easy to calculate the interference following the same approach, but it would need a treatment of the attenuation which is beyond the scope of this article. In any case there are two possibilities for the shower types: (i) An electromagnetic shower and a hadronic shower, with the electromagnetic shower in the electronic decay of the $\tau$, and the hadronic shower induced by the products of the collision with the nucleon; (ii) two hadronic showers when the $\tau$ decays hadronically.
\end{enumerate} 

For practical purposes it is sufficient to simulate purely electromagnetic showers and purely hadronic showers at different energies, as will be shown in the following. A combination of these two will be shown to be sufficient to accurately describe the radio pulses induced by any of the three types of showers described above. 

\section{Results}
\label{S:Results}
\subsection{Electromagnetic and hadronic form factors}

The time-domain vector potential $R A(\theta_C,t)$ at the Cherenkov angle for an observer in the Fraunhofer region of the shower at a distance $R$ is obtained with the ZHAireS Monte Carlo code. This allows us to obtain the form factor associated to the lateral spread of the shower with the aid of Eq.\,(\ref{eq:lambda}). 

\begin{figure}[ht]
{\centering 
\resizebox*{0.51\textwidth}{!}{\includegraphics{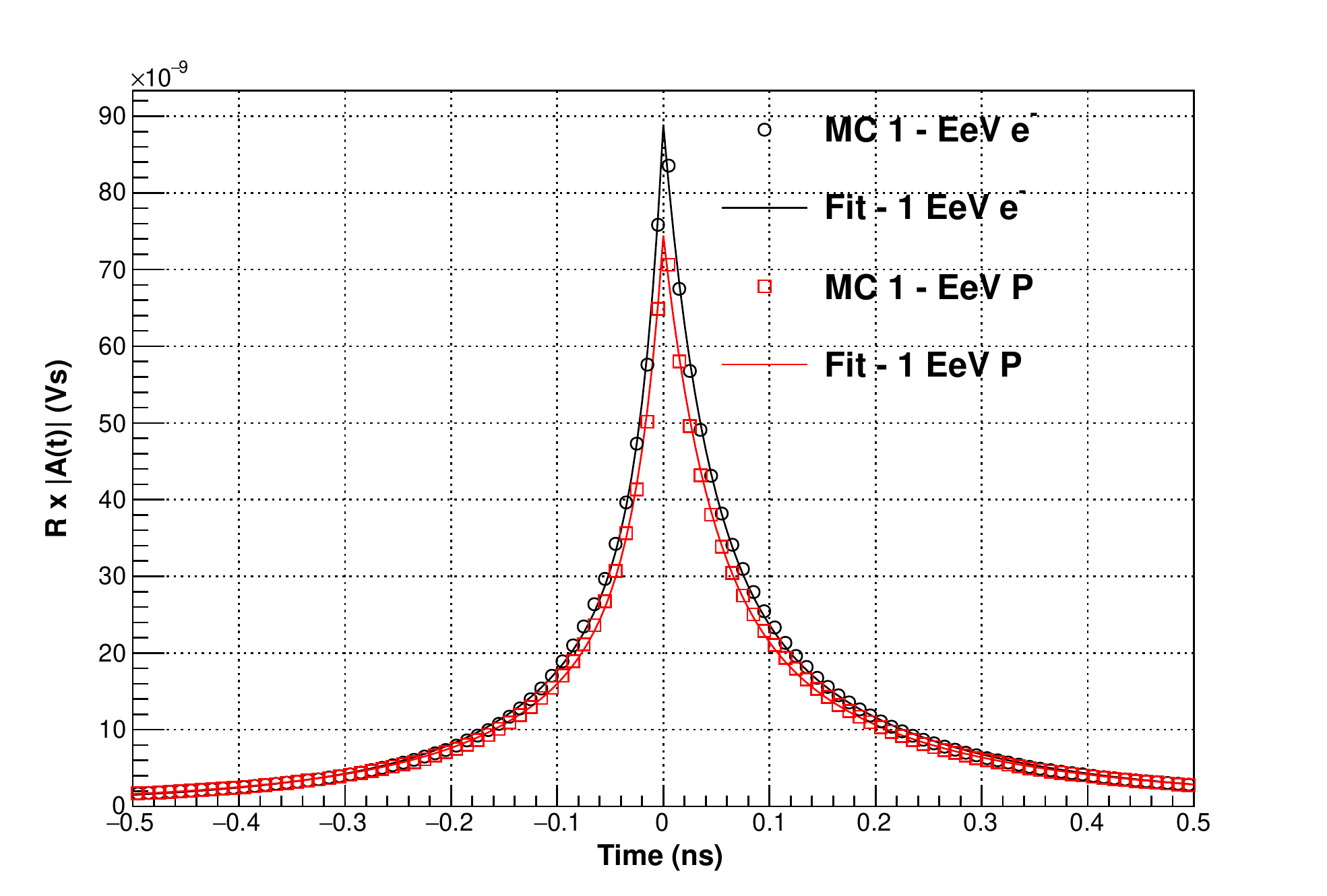}}
\par}
\caption{The vector potential at the Cherenkov angle obtained in ZHAireS simulations of showers induced by a 1 EeV electron (black dots) and a 1 EeV proton (red dots) in homogeneous ice (density $\rho$ = 0.924 g cm$^{-3}$ and refractive index n=1.78, Cherenkov angle $\theta_{C} \sim 55.8^{0}$). 
The vector potential obtained in the ZHAireS simulations is compared to that given by $A_{\rm EM}(1\,{\rm EeV},\theta_C,t)$ in Eq.\,(\ref{eq:vp_fit_EM}) (solid black line) and $A_{\rm HAD}(1\,{\rm EeV},\theta_C,t)$ in Eq.\,(\ref{eq:vp_fit_HAD}) (solid red line).}
\label{fig:vp_EM_vs_HAD}
\end{figure}

Electromagnetic showers initiated by an electron, and hadronic showers initiated by a proton were simulated for this purpose. The vector potentials were parameterized, and with the aid of Eqs.\,(\ref{eq:model_Fres}) and (\ref{eq:lambda}) the vector potential for an arbitrary observer in the radiative field region of the shower can be obtained. The electromagnetic $A_{\rm EM}(\theta_C,t)$ and hadronic $A_{\rm HAD}(\theta_C,t)$ vector potentials are shown in Fig.~\ref{fig:vp_EM_vs_HAD}. The shape of the potentials is slightly different, reflecting the differences in the lateral distribution function and the radial dependence of the particle velocity in the two types of showers. It is interesting to see that they are both not symmetric around $t=0$, due to the radial components of the velocity of the particles pointing in different directions as seen by an observer at azimuth $\phi=0$. Also, the electromagnetic vector potential is slightly higher than the hadronic one. This is due to the so-called missing energy mainly carried by muons and neutrinos in hadronic showers, and not contributing significantly to the excess charge and hence to the production of radio waves \cite{ZHAireS_ice}. 

The electromagnetic $A_{\rm EM}(\theta_C,t)$ vector potential for an electron of energy $E_e$ was parameterized  with the following function:
\begin{equation}
\begin{split}
&
R\times A_{\rm EM}(E_e,\theta_C,t) = -4.445\times 10^{-8}~[\mbox{V s}] 
\times~\frac{E_e}{1\,{\rm EeV}}
\\
&
\left\{ 
\begin{array}{l l}
\exp\left(-\frac{|t|}{0.0348}\right)+(1+2.298|t|)^{-3.588} & \quad \mbox{if ~$t>0$}\\ \\
\exp\left(-\frac{|t|}{0.0203}\right)+(1+2.616|t|)^{-4.043} & \quad \mbox{if ~$t<0$}\,. \\ 
\end{array} 
\right. 
\end{split}
\label{eq:vp_fit_EM}
\end{equation}
This parameterization updates the one given in \cite{ARZ11}, and it is significantly more accurate in the dominant part of the peak (see Fig.~\ref{fig:vp_EM_vs_HAD}).

In the case of a hadronic shower initiated by a proton of energy $E$, the vector potential $A_{\rm HAD}(\theta_C,t)$ obtained in ZHAireS simulations can be parameterized as,
\begin{equation}
\begin{split}
&
R\times A_{\rm HAD}(E,\theta_C,t) = -4.071\times 10^{-8}~[\mbox{V s}] ~ 
\times \frac{E_{\rm em}(E)}{1\,{\rm EeV}}
\\
&
\left\{ 
\begin{array}{l l}
\exp\left(-\frac{|t|}{0.0391}\right)+(1+2.338|t|)^{-3.320} & \quad \mbox{if ~$t>0$}\\ \\
\exp\left(-\frac{|t|}{0.0234}\right)+(1+2.686|t|)^{-3.687} & \quad \mbox{if ~$t<0$}\,. \\ 
\end{array} 
\right. 
\end{split}
\label{eq:vp_fit_HAD}
\end{equation}
%
Here, $E_{\rm em}$ represents the energy transferred to the electromagnetic component of the shower, responsible for the bulk of the radio emission due to the excess negative charge, and which depends on the energy of the primary particle $E$. Performing Monte Carlo simulations of proton showers in ice at different energies with ZHAireS we have obtained a phenomenological parameterization for $E_{\rm em}(E)$:
\begin{equation}
\begin{split}
&
E_{\rm em}(E) = f(\epsilon) \times E = 
\\ 
&
(-21.98905-2.32492\,\epsilon+0.019650\,\epsilon^2+13.76152\,\sqrt\epsilon) \times E 
\\ 
\nonumber
\end{split}
\end{equation}
with $\epsilon=\log_{10}(E/{\rm eV})$. The electromagnetic energy $E_{\rm em}$ does not exactly scale linearly with the
hadronic energy $E$. The reason is that the fraction $f$ of the hadronic energy that is lost to neutrinos 
and muons also depends on $E$ \cite{InvisibleEnergy}. 
The two functions in Eqs.\,(\ref{eq:vp_fit_EM}) and (\ref{eq:vp_fit_HAD}) are shown in Fig.\,\ref{fig:vp_EM_vs_HAD} compared to the result of the corresponding full simulation with ZHAireS. The agreement between the fits and the simulations is at the $\pm 3\%$ level in the time range $(-0.5,0.5)$ ns in which the numerical value of the vector potential falls by two orders of magnitude with respect to the value at the peak. 

\begin{figure}[ht]
{\centering 
\resizebox*{0.51\textwidth}{!}{\includegraphics{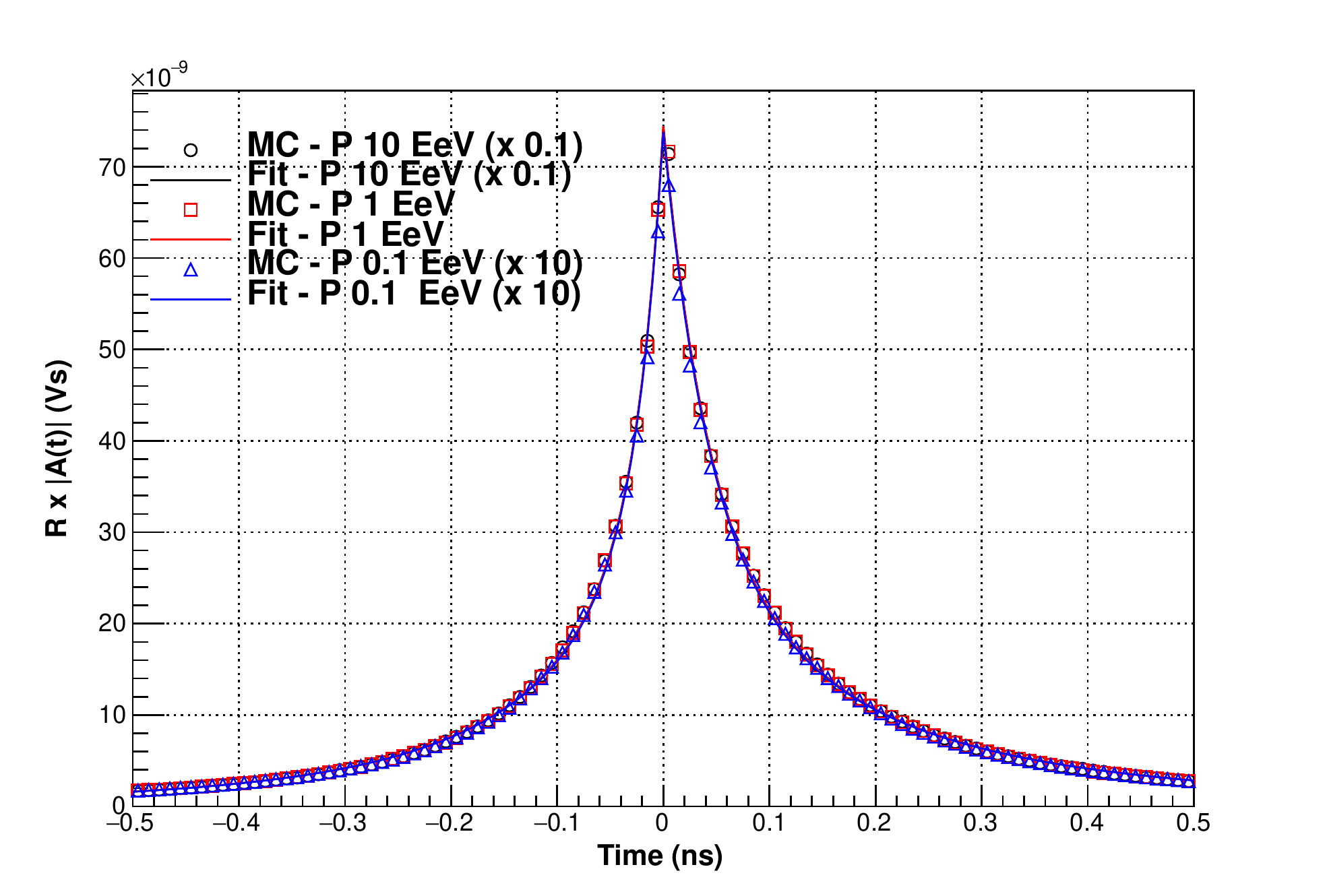}}
\par}
\caption{The vector potential at the Cherenkov angle obtained in ZHAireS simulations of showers induced by protons of different energies, $E_p=$0.1 EeV (blue dots, scaled up by a factor 10), $E_p=$1 EeV (red dots) and $E_p=$10 EeV (black dots, scaled down by a factor 0.1) in homogeneous ice.  
The vector potential obtained in the ZHAireS simulations is compared to those given by $A_{\rm HAD}(E_p,\theta_C,t)$ in Eq.\,(\ref{eq:vp_fit_HAD}) at the corresponding proton energy (solid blue, red and black lines).
}
\label{fig:vp_HAD_energy}
\end{figure}

In Fig.\,\ref{fig:vp_HAD_energy} we explore the energy behavior of the parameterization in Eq.\,(\ref{eq:vp_fit_HAD}) by comparing the vector potential obtained in ZHAireS simulations of proton-induced showers at different energies. At the energies of interest in this work, the differences between the fit and the ZHAireS simulations continue to be on the order of $\pm\,5\%$ in the time interval $(-1,1)$ ns. 

With the electromagnetic and hadronic vector potentials given respectively in Eqs.~(\ref{eq:vp_fit_EM}) and (\ref{eq:vp_fit_HAD}), we can construct the vector potential for all the neutrino flavors and interaction channels described in Section \ref{S:First}:

\begin{enumerate}

\item 
For mixed showers produced in $\nu_e$ CC interactions at energy $E_\nu$, the vector potential can be obtained as a sum of the electromagnetic and hadronic vector potentials at the energy carried by each shower in the interaction, namely,
\begin{equation}
\begin{split}
&
R\times A_{\nu_e}^{\rm CC}(E_\nu,\theta_C,t) = 
\\
&
R\times A_{\rm EM}[(1-y)E_\nu,\theta_C,t] ~ + 
\\ 
&
R\times A_{\rm HAD}(yE_\nu,\theta_C,t),
\end{split}
\label{eq:vp_nue_CC}
\end{equation}
with $A_{\rm EM}$ and $A_{\rm HAD}$ given in Eqs.\,(\ref{eq:vp_fit_EM}) and (\ref{eq:vp_fit_HAD}) respectively. An example is shown in the top panel of Fig.\,\ref{fig:vp_nu}, where we compare the vector potential as obtained with Eq.\,(\ref{eq:vp_nue_CC}) for different values of the fraction of neutrino energy transferred to the nucleus, $y$, with those obtained directly in full ZHAireS simulations of $\nu_e$-induced showers in ice. The difference between the fit and the ZHAireS simulations is at the $\pm\,3\%$ level in the time range spanning the dominant part of the peak. This confirms the validity of this approach for the calculation of the vector potentials in $\nu$-induced showers. As expected, the peak value of $A_{\nu_e}^{\rm CC}$ is roughly independent of the value of $y$, reflecting that most of the neutrino energy goes into the mixed (electromagnetic + hadronic) shower regardless of the value of $y$. However, the shape of the vector potential is slightly dependent on $y$, being increasingly wider as the hadronic component of the mixed shower increases, i.e. as $y$ increases (see also Fig.\,\ref{fig:vp_EM_vs_HAD}).

\item
For hadronic showers produced by the ``jet" in $\nu_X$ NC, or by $\nu_\mu$ and $\nu_\tau$ CC interactions, the vector potential can be obtained as:
\begin{equation}
R\times A_{\nu}(E_\nu,\theta_C,t) = 
R\times A_{\rm HAD}(yE_\nu,\theta_C,t).
\label{eq:vp_nu_NC}
\end{equation}
with $A_{\rm HAD}$ given in Eq.\,(\ref{eq:vp_fit_HAD}).

\begin{figure}[ht]
{\centering 
\begin{tabular}{c}
\resizebox*{0.53\textwidth}{!}{\includegraphics{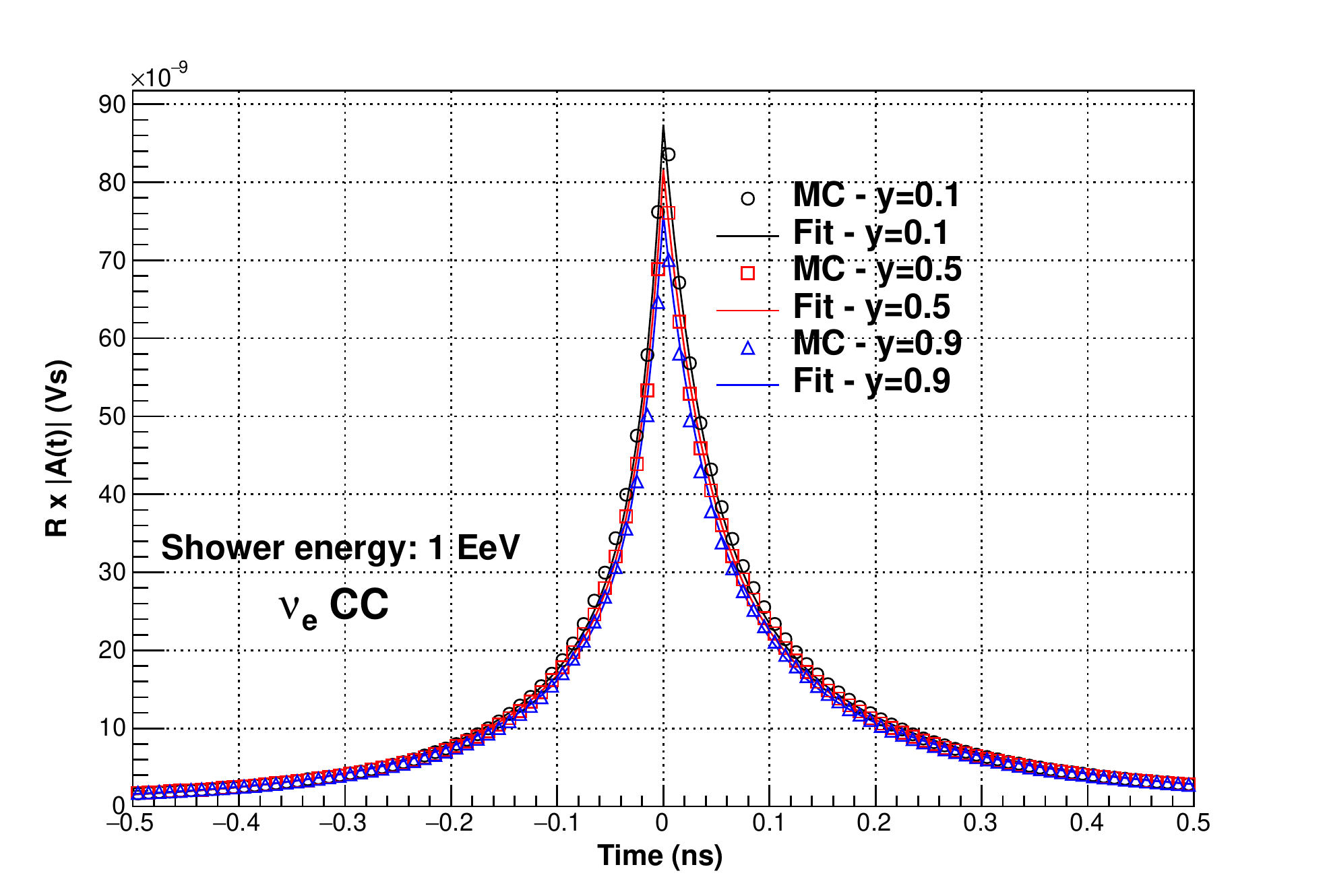}} \\
\resizebox*{0.53\textwidth}{!}{\includegraphics{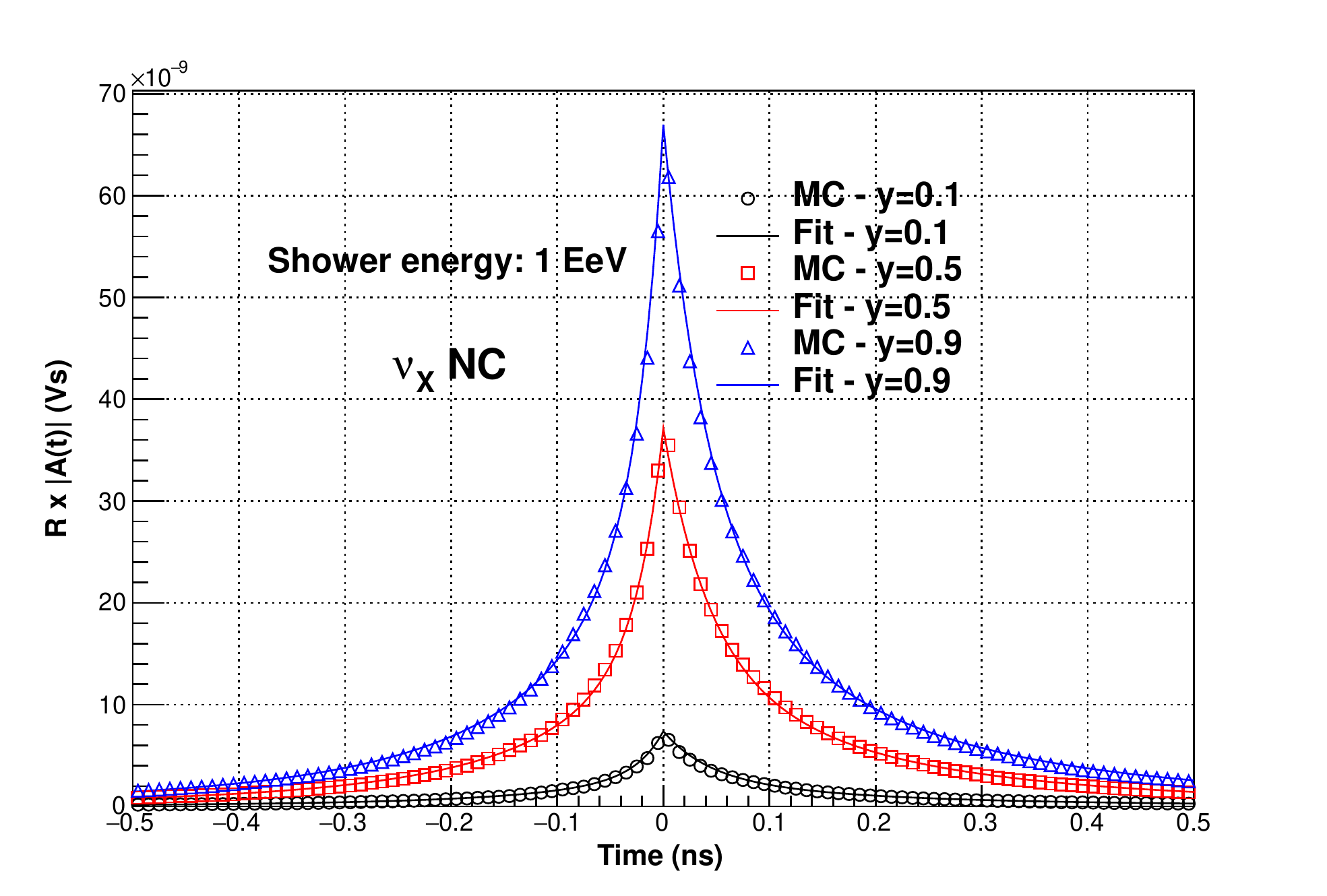}}
\\
\end{tabular}\par}
\caption{The vector potential at the Cherenkov angle (filled dots) obtained in ZHAireS simulations of showers induced by a $\nu_e$ in CC interactions (top panel), and by the products of a NC $\nu$ interaction (bottom panel) in homogeneous ice,  
with the numbers in parenthesis indicating the branching fractions of each decay channel. In all cases, the neutrino energy is $E_\nu=1$ EeV, and we show the vector potential for values of the fractional energy transferred to the nucleus $y=0.1$, $y=0.5$ and $y=0.9$. The vector potentials obtained in ZHAireS simulations are compared to those given in Eq.\,(\ref{eq:vp_nue_CC}) and Eq.\,(\ref{eq:vp_nu_NC}) (solid lines).
}
\label{fig:vp_nu}
\end{figure}

An example is shown in the bottom panel of Fig.\,\ref{fig:vp_nu}, where we compare the vector potential as obtained with Eq.\,(\ref{eq:vp_nu_NC}) for different values of the fractional energy transfer $y$, with those obtained directly in full ZHAireS simulations of $\nu$-induced showers in ice. The agreement between Eq.\,(\ref{eq:vp_nu_NC}) and the simulations is at the few percent level.

\item
The parameterizations given in Eqs.\,(\ref{eq:vp_fit_EM}) and (\ref{eq:vp_fit_HAD}) also allow us to obtain the vector potential produced in the shower initiated by the decay of the $\tau$ lepton in a $\nu_\tau$ CC interaction. They reflect the two main types of showers induced, namely, electromagnetic and hadronic. They will be assumed to be universal in the sense that the effective lateral distribution function is expected to be the same independently of shower energy or on whether the showers are initiated in the neutrino vertex, in the hadron vertex or in the tau decay. 

For the electronic decay channel, denoting as $f_e$ the fraction of the tau-lepton energy carried by the electron, the vector potential is simply given by,
\begin{equation}
R\times A_{\tau}^e(E_\tau,\theta_C,t) = R\times A_{\rm EM}(f_e E_\tau,\theta_C,t).
\label{eq:vp_tau_e}
\end{equation}
For the $\tau$ decay into hadrons, denoting $f_h$ the fraction of energy carried by hadrons in the decay, the vector potential can be obtained as,
\begin{equation}
R\times A_{\tau}^{\rm had}(E_\tau,\theta_C,t) = R\times A_{\rm HAD}(f_h E_\tau,\theta_C,t).
\label{eq:vp_tau_hadrons}
\end{equation}
The number and energy distribution of hadrons in a proton interaction and the corresponding distribution of hadrons in a $\tau$ decay differ significantly. However, using the vector potential in Eq.\,(\ref{eq:vp_fit_HAD}) to model the hadronic decays of the $\tau$ turns out to be a good approximation, regardless of the hadronic decay channel. This is shown in  Fig.\,\ref{fig:vp_nutau}. For this purpose we have simulated with ZHAireS showers induced by the secondaries produced in the three most frequent hadronic decay channels of the $\tau$ lepton, namely: 
(i) $\tau^-\rightarrow\pi^- + \nu_\tau$ ($\sim\,10.8\%$); 
(ii) $\tau^-\rightarrow\pi^- + \pi^0 + \nu_\tau$ ($\sim\,25.5\%$), and 
(iii) $\tau^-\rightarrow\pi^- + \pi^0 + \pi^0 + \nu_\tau$  ($\sim\,9.3\%$), 
where the numbers in parenthesis indicate the corresponding branching fractions. 
In these three cases, the sum of the energy of the secondaries excluding the $\nu_\tau$, i.e. the shower energy, was fixed to 0.5 EeV. The vector potential obtained in ZHAireS simulations is compared to that obtained using the approximation in Eq.\,(\ref{eq:vp_tau_hadrons}) with agreement again at the few percent level.
As can also be seen in Fig.\,\ref{fig:vp_nutau}, there is a very weak dependence of the vector potential on the hadronic decay channel of the $\tau$. 

\begin{figure}[ht]
{\centering 
\resizebox*{0.53\textwidth}{!}{\includegraphics{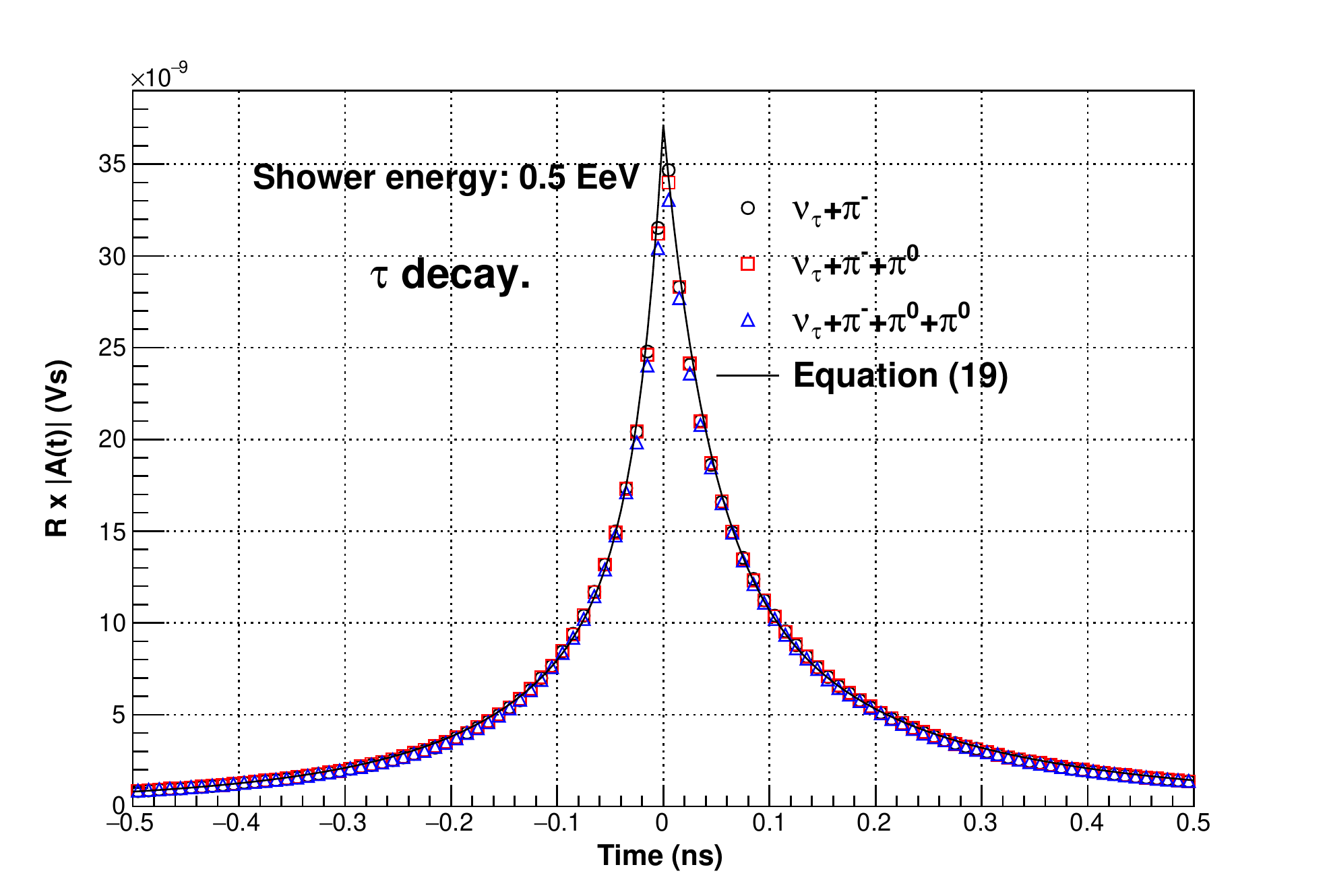}}
\par}
\caption{The vector potential at the Cherenkov angle (filled dots) obtained in ZHAireS simulations of showers induced by the decay products of a $\tau$ in homogeneous ice. 
We show the vector potential of showers induced by the secondaries produced in the three most frequent hadronic decay channels: $\tau^-\rightarrow\pi^- + \nu_\tau$ ($\sim 10.8\%$) , $\tau^-\rightarrow\pi^- + \pi^0 + \nu_\tau$ ($\sim 25.5\%$), and $\tau^-\rightarrow\pi^- + \pi^0 + \pi^0 + \nu_\tau$  ($\sim 9.3\%$). The sum of the energy of the secondaries excluding the $\nu_\tau$, i.e. the shower energy, was chosen to be 0.5 EeV in all cases for the purposes of comparison. The vector potential obtained in ZHAireS simulations is compared to that obtained using the approximation in Eq.\,(\ref{eq:vp_tau_hadrons}) (solid line).
}
\label{fig:vp_nutau}
\end{figure}

\end{enumerate}

\subsection{Full algorithm for the calculation of Askaryan emission}
\label{SS:algorithm}

We have applied the following computational algorithm to obtain the Askaryan emission in neutrino-induced showers for an arbitrary observer: 

\begin{enumerate}

\item 
Obtain the specific form factor 
combining the electromagnetic and hadronic form factors  respectively given by the parameterizations,
Eqs.\,(\ref{eq:vp_fit_EM}) and (\ref{eq:vp_fit_HAD}), as follows:  

\begin{enumerate}
\item For a $\nu_e$ CC interaction Eq.\,(\ref{eq:vp_nue_CC}) should be used. 
\item For the shower induced by the secondary products of the struck nucleon in a $\nu_\tau$ CC interaction, Eq.\,(\ref{eq:vp_nu_NC}) 
can be used again.
\item For a $\nu$ of any flavor in a NC interaction, the vector potential is obtained with Eq.\,(\ref{eq:vp_nu_NC}). 
\item For a $\tau$ produced in a $\nu_\tau$ CC interaction and decaying hadronically one can obtain the vector potential 
with Eq.~(\ref{eq:vp_tau_hadrons}). If the $\tau$ decays in an electron, Eq.\,(\ref{eq:vp_tau_e}) should be used.  
\end{enumerate}

\item 
Obtain the charge excess longitudinal profile of the neutrino-induced shower~$Q(z')$. This can be provided as either the output of a particle shower simulation such as ZHAireS or from a parameterization. 

\item
The vector potential obtained in item 1 along with the total charged track-length  
$LQ_{tot}=\int dz'Q(z')$ (obtained with a direct integration of the longitudinal profile of the excess charge), 
allows to extract the functional form of the form factor $F_p$ to be used in Eq.\,(\ref{eq:model_Fres}).

\item
Convolve $F_p$ with $Q(z')$ according to Eq.\,(\ref{eq:model_Fres}) to obtain the time-domain vector potential 
for an arbitrary observer. 

\item 
The electric field at the observer's position is simply obtained from a numerical derivative of the vector potential with respect to time:
${\mathbf{E}}=-\partial {\mathbf{A}}/\partial t$. 

\end{enumerate}
%

\subsection{Askaryan emission in neutrino-induced showers}

We have applied the algorithm in Section \ref{SS:algorithm} to obtain the vector potential induced by the different types of neutrino-induced showers enumerated in Section \ref{S:TheSimulations}, and in the following we report the comparison with the vector potential obtained in full ZHAireS simulations of the same showers.

As explained in detail in \cite{ARZ11}, the time dependence of the vector potential traces the change of the excess charge as the shower develops through the retarded time $t'$ \cite{Feynman_1963}. At any instant $t$, there is a retarded time at which the main contribution to the vector potential takes place, and as time passes its normalization traces that of the longitudinal profile of the charge excess, reproducing its features such as peaks. There is however a compression factor to change from the retarded to the observer time intervals which depends on the observation angle relative to the shower axis. The compression factor diverges for observers at the Cherenkov angle, so that they see all the shower development at once, except for the tails of the form  factor due to the lateral spread. In this case, the time dependence of the vector potential is that of the form factor function. For observers away from the Cherenkov angle the compression factor is smaller so that the pulse traces the longitudinal development of the charge. Also observers outside the Cherenkov cone see the beginning of the shower first while those inside the Cherenkov angle angle see it in reverse order. These general features of the vector potential are present in all the case examples depicted below. 

\begin{figure}[h]
{\centering 
\resizebox*{0.48\textwidth}{!}{\includegraphics{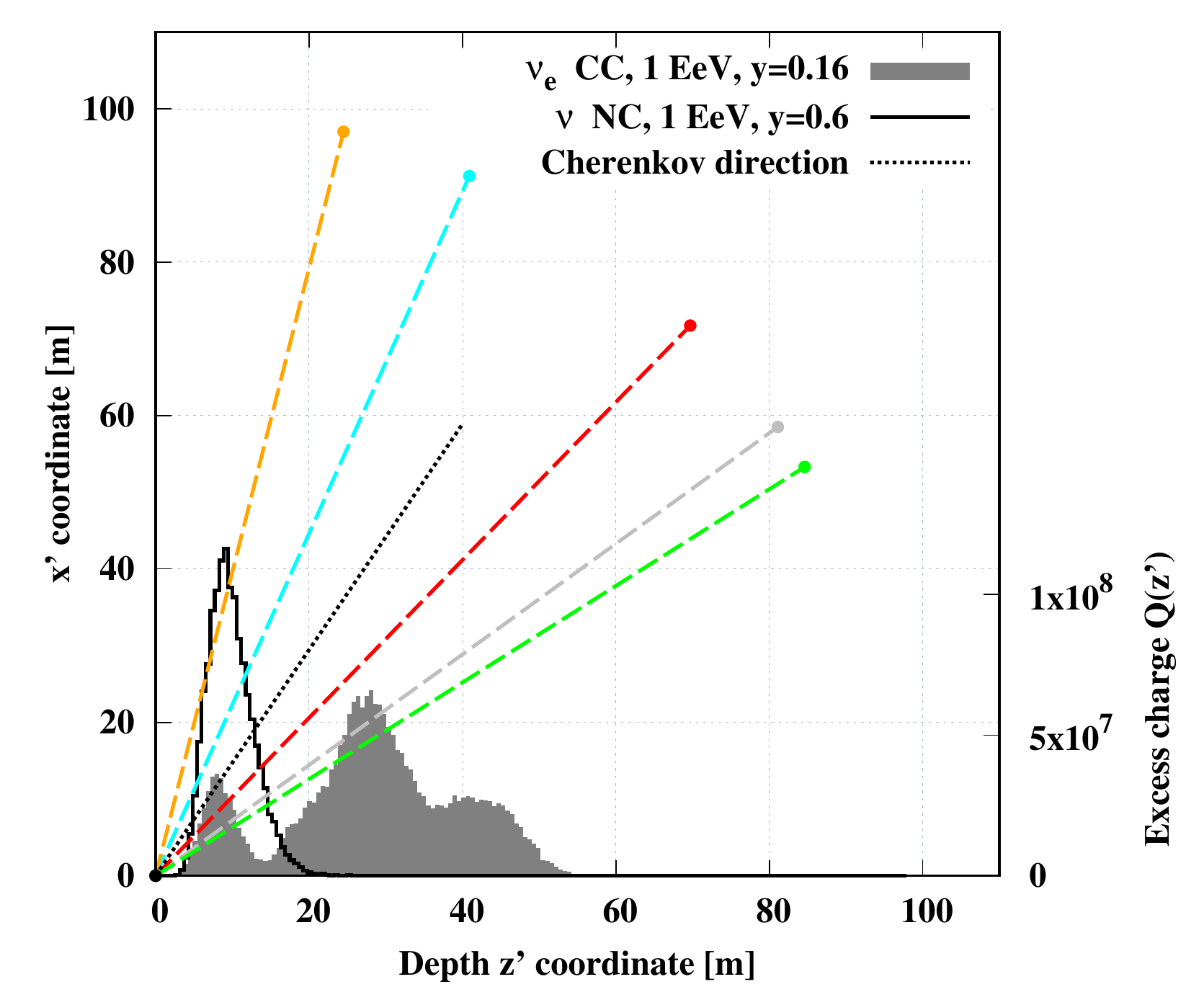}}
\par}
\caption{Positions of observers where the vector potential, depicted in Figs.~\ref{fig:vp_nue_CC_y0p2}, \ref{fig:vp_nue_CC_nu_NC_y0p1}, \ref{fig:vp_nue_CC_NC_yvar} and \ref{fig:vp_tau_decay}, is calculated. The color-coding follows the same as in the figures. The positions are given in coordinates $z'$ (along shower axis) and $x'$ (perpendicular to shower axis - left scale of the plot) relative to the starting point of the shower. The dotted black line indicates the Cherenkov direction, for reference. The longitudinal profile of the excess charge $Q(z')$ of two neutrino-induced showers is also shown (right scale of the plot).
}
\label{fig:observers}
\end{figure}

\begin{figure}[h]
{\centering 
\resizebox*{0.51\textwidth}{!}{\includegraphics{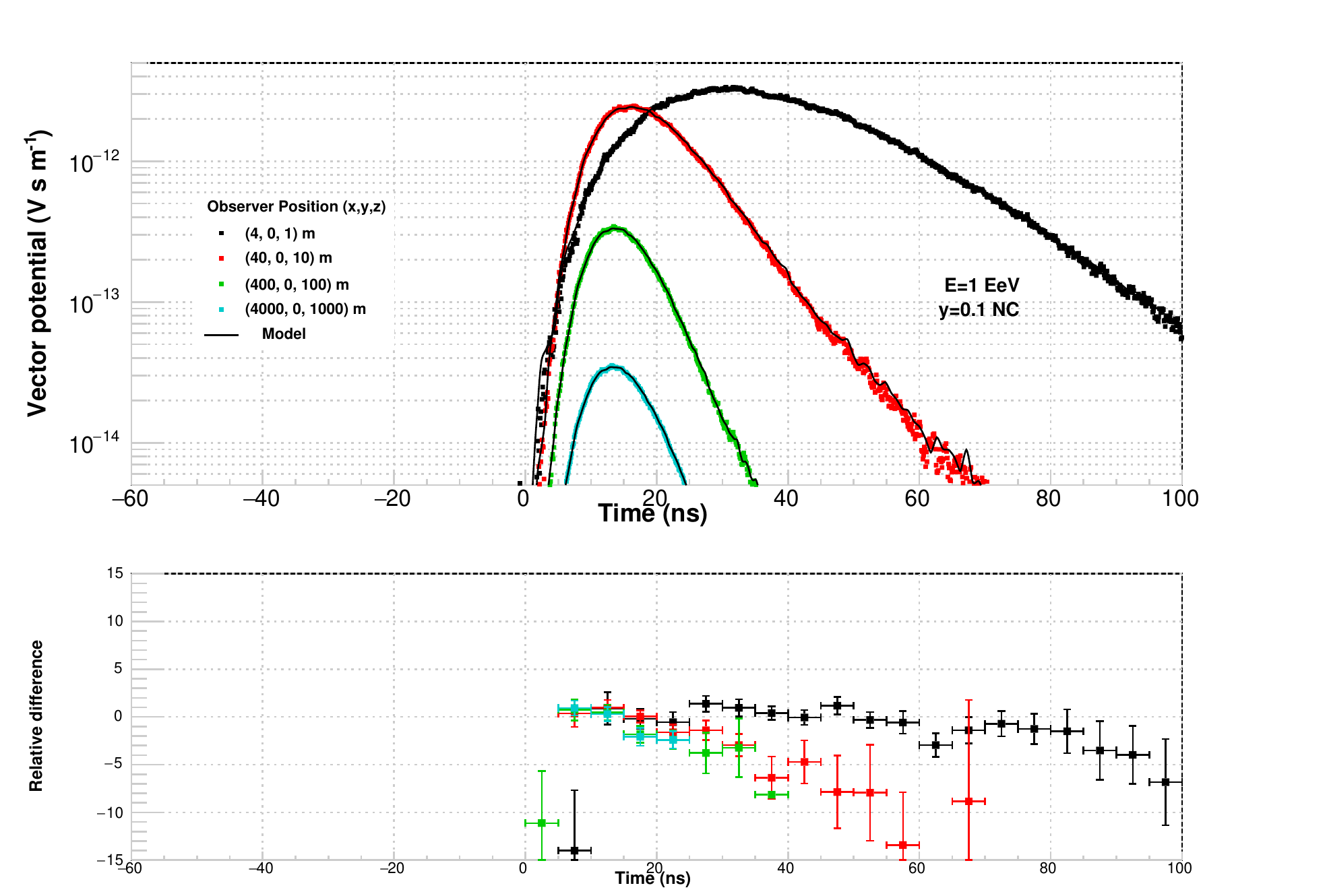}}
\par}
\caption{The vector potential in showers in homogeneous ice induced by a $\nu$ NC interaction. The neutrino energy is $E_\nu=$1 EeV and the energy fraction carried by the secondaries in the hadronic vertex of the  interaction is $y=0.1$. The solid black lines represent the vector potentials as obtained with our model, applying the algorithm explained in Section \ref{SS:algorithm}, while the colored lines correspond to the vector potential obtained in full ZHAireS simulations of the same showers. The vector potential was obtained for observers at six distances with respect to the starting point of the shower, from $\sim\,4.13$ m to $\sim\,4.13$ km in steps of factors of 10, but at the same angle ($\sim 75^\circ$) with respect to the shower axis. 
For each vector potential we also show the relative difference between the full ZHAireS simulations and the approach in this article, normalized to the ZHAireS results and averaged in bins of 5 ns width. The rms value of the average difference in each bin is also shown. The accuracy of the model worsens at times when the amplitude of the vector potential also drops significantly with respect to the value at the peak and is not expected to be relevant for practical purposes.
}
\label{fig:vp_nu_NC_distance}
\end{figure}

We now compare the pulses obtained with our method for different types of neutrino interactions to those obtained in full MC simulations with ZHAireS, to test the fidelity of the model and explore its performance as the energy, the fractional energy transfer and observation angle and distance are changed. In each set of comparisons (but the one in Fig.\,\ref{fig:vp_nu_NC_distance}) we will always consider five observer positions, all at a distance of 100 m from the start of the shower. The positions can be seen in the diagram in Fig.~\ref{fig:observers} in relation to two shower profiles, one of them (shaded) with characteristic multiple peaks due to the LPM effect. 
One observer (red line in Fig.~\ref{fig:observers}) is chosen making an angle of $\sim45^\circ$, close to the Cherenkov direction as established from the shower maximum of a typical hadronic shower (without significant elongation due to the LPM effect). Two other positions are chosen at larger angles (blue and orange lines in all figures) and two more positions at smaller angles (grey and green lines). The vector potentials for observers at positions at the lowest angles with respect to the starting point of the shower (red, grey and green), i.e. within the Cherenkov cone, display the  characteristic time inversion and arrive later in contrast to the vector potential from the positions with larger angles (cyan and orange) where observers are placed outside the Cherenkov cone. 

\begin{figure}[h]
{\centering 
\begin{tabular}{c}
\resizebox*{0.51\textwidth}{!}{\includegraphics{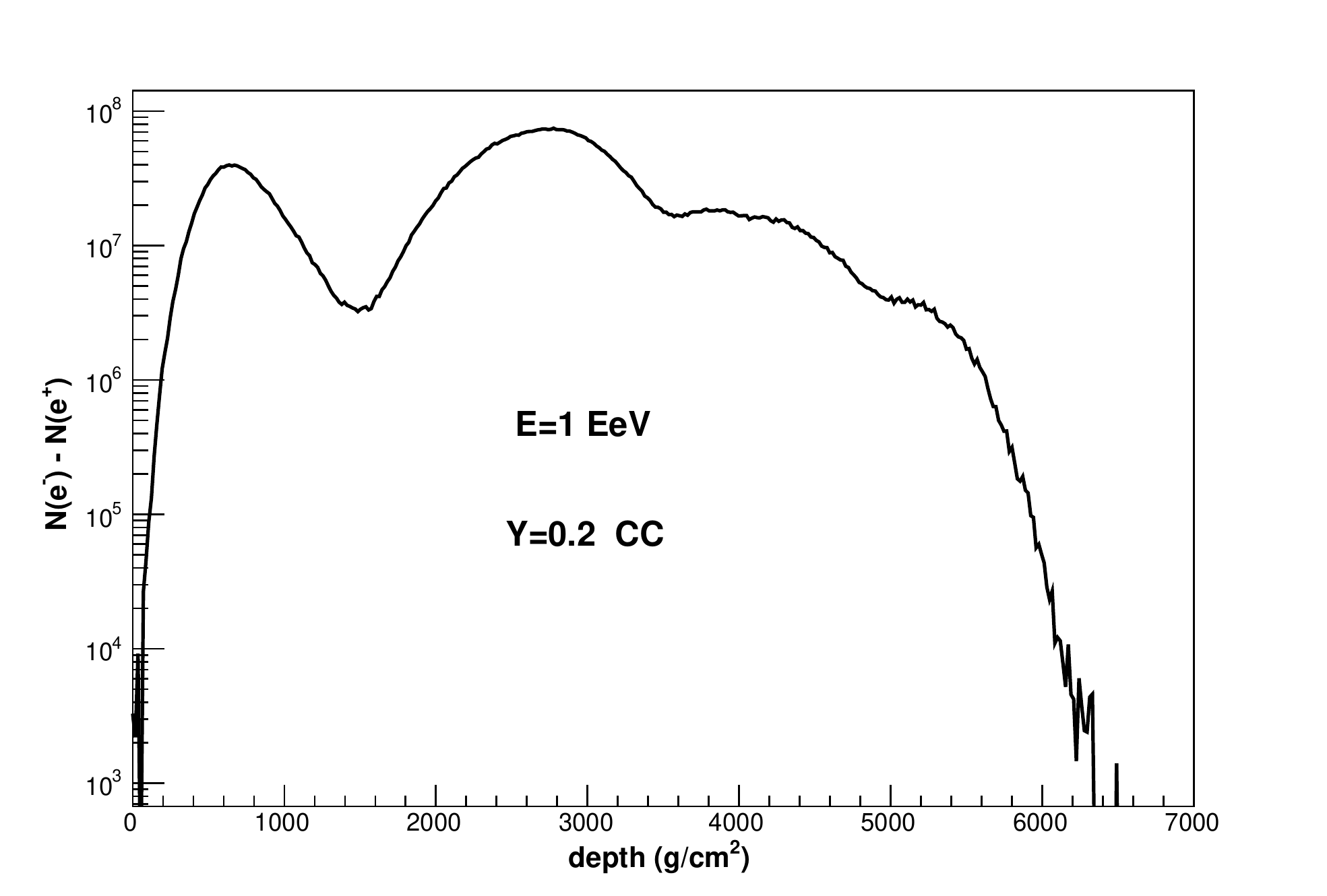}} \\
\resizebox*{0.51\textwidth}{!}{\includegraphics{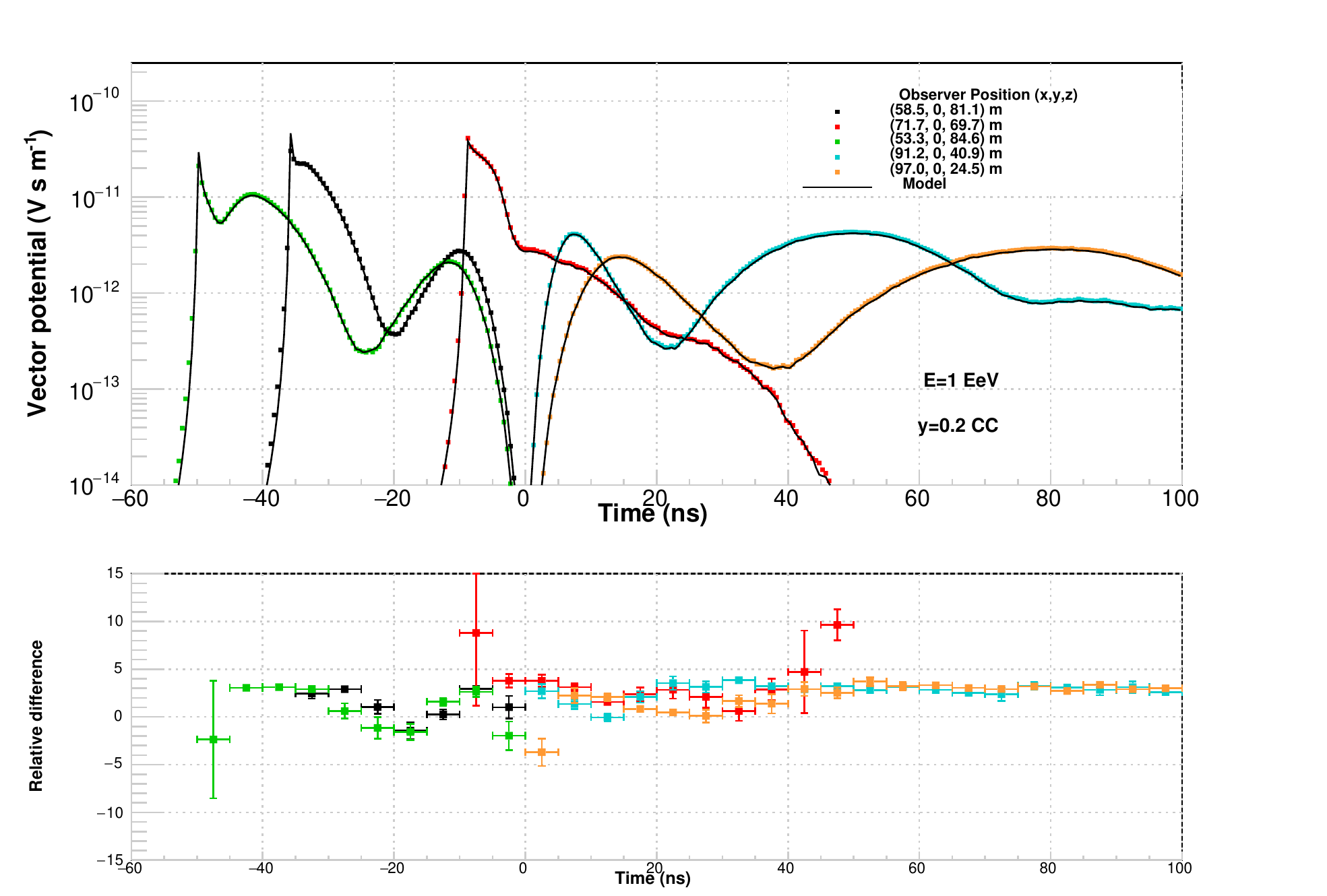}} \\
\resizebox*{0.51\textwidth}{!}{\includegraphics{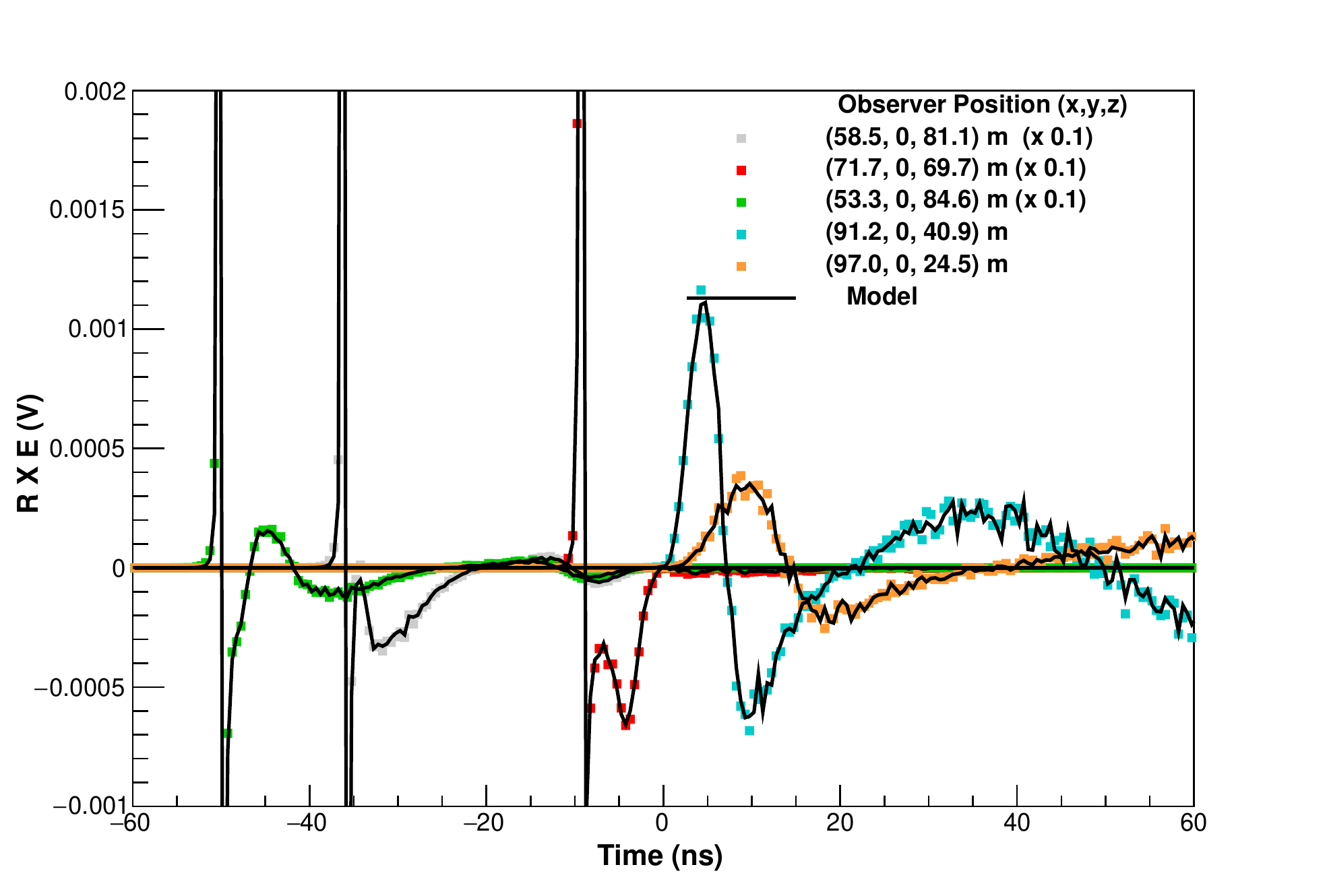}} \\
\end{tabular}\par}
\caption{Top panel: Longitudinal profile of the excess charge of a shower in homogeneous ice, induced in a $\nu_e$ charged-current interaction with $E_\nu=1$ EeV and $y=0.2$. Second panel from the top: The vector potential due to the same shower. Bottom panel: The associated electric field. In both the second and bottom panels, the solid black lines are the values obtained with the approach of this article~(Section \ref{SS:algorithm}) while the colored dots correspond to the values directly obtained in full ZHAireS simulations of the same showers.  Each curve corresponds to an observer located 100~m away from the starting point of the shower at a different location (see diagram in Fig.~\ref{fig:observers}). Third panel from the top: Relative difference between the full ZHAireS simulations and the approach in this article, normalized to the ZHAireS results and averaged in bins of 5 ns width. The rms value of the average difference in each bin is also shown. 
}
\label{fig:vp_nue_CC_y0p2}
\end{figure}

As a first example we show in Fig.\,\ref{fig:vp_nue_CC_y0p2} the vector potential (middle panel) and electric field (bottom panel) for a shower in homogeneous ice induced by $\nu_e$ CC interaction with $E_\nu=1$ EeV and a fraction of energy transferred to the struck nucleon $y=0.2$. The shower exhibits a multi-peaked structure in the longitudinal profile of the excess charge, due to the LPM effect as also shown in Fig.\,\ref{fig:vp_nue_CC_y0p2} (top panel).  The vector potential obtained with our approach agrees with that of the simulations to order $3\%$ in the region around the peak of the pulse. Due to the smaller accuracy of the fits to the vector potentials given in Eqs.\,(\ref{eq:vp_fit_EM}) and (\ref{eq:vp_fit_HAD}) in the time region outside the $(-0.5,0.5)$ ns window, the accuracy between the model and the full MC simulations drops to about $20\%$ in the onset and in the latest times of the pulse. However, the amplitude also drops by at least three orders of magnitude in these regions and it is not expected to be relevant for practical purposes. The accuracy of our methodology is easier to view in the vector potential that can be plotted in logarithmic scale, in contrast to the electric field. For this reason, we will only show the vector potential in the plots that follow.

\begin{figure*}[h]
{\centering 
\begin{tabular}{cc}
\resizebox*{0.51\textwidth}{!}{\includegraphics{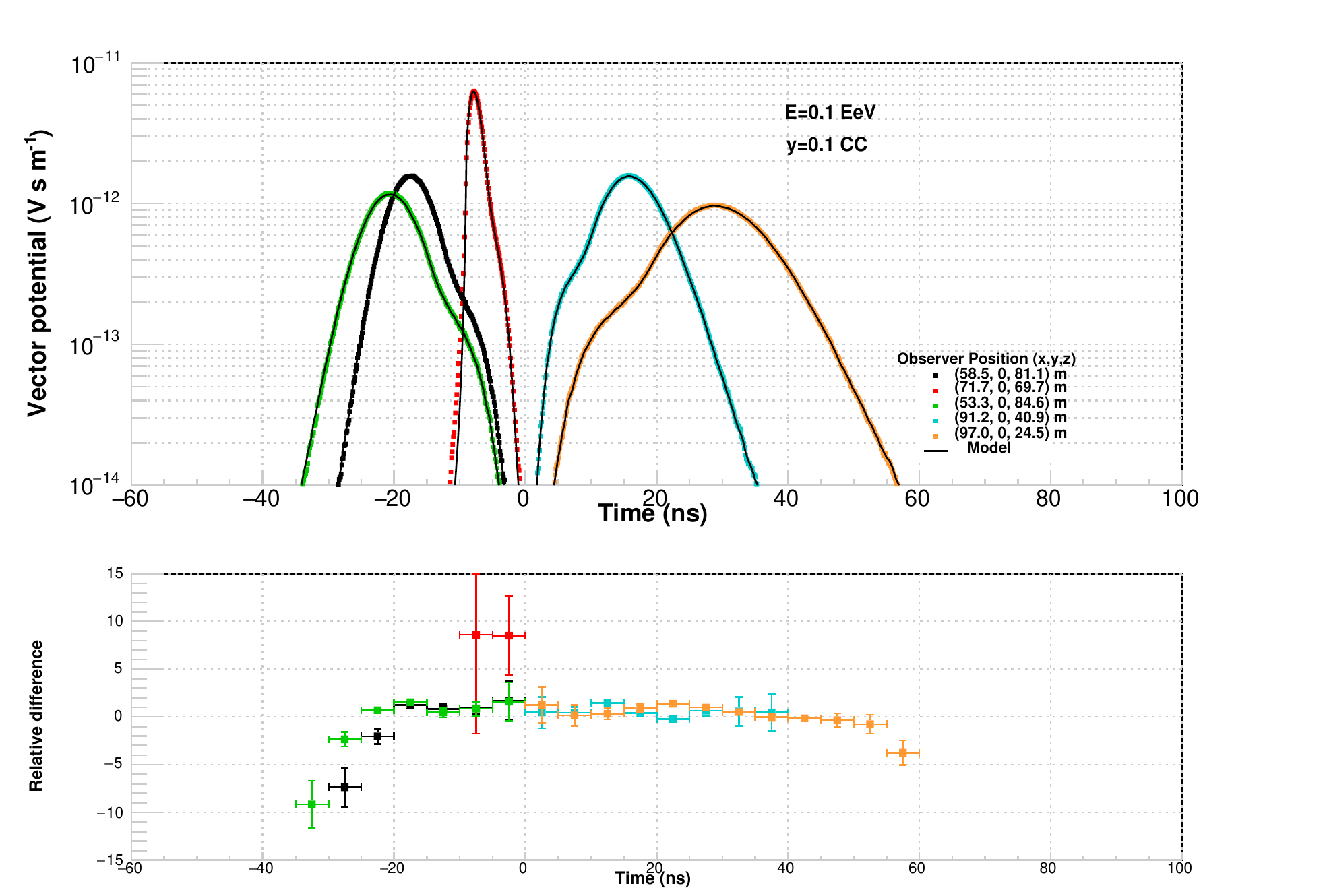}}
&\resizebox*{0.51\textwidth}{!}{\includegraphics{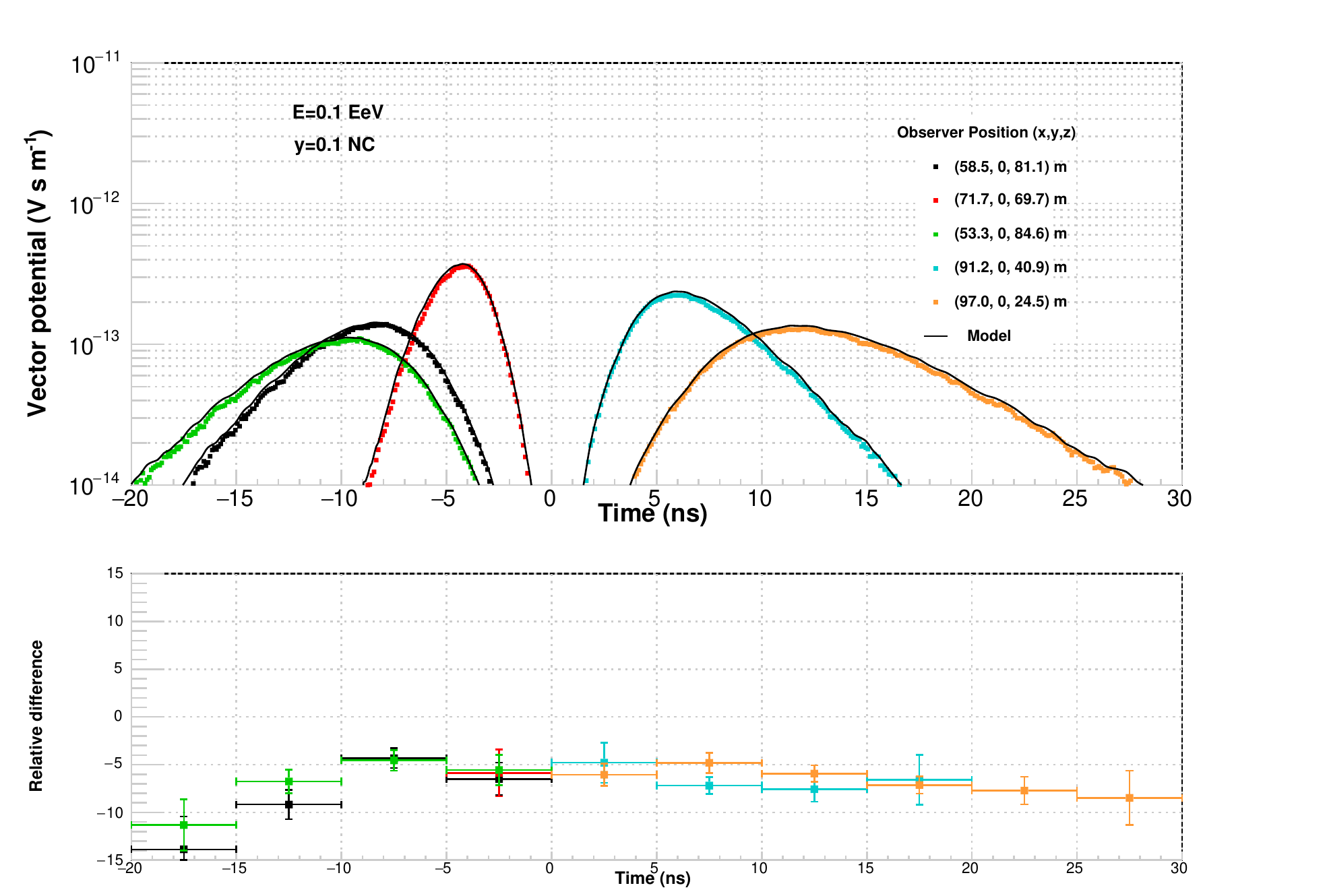}}
\\
\resizebox*{0.51\textwidth}{!}{\includegraphics{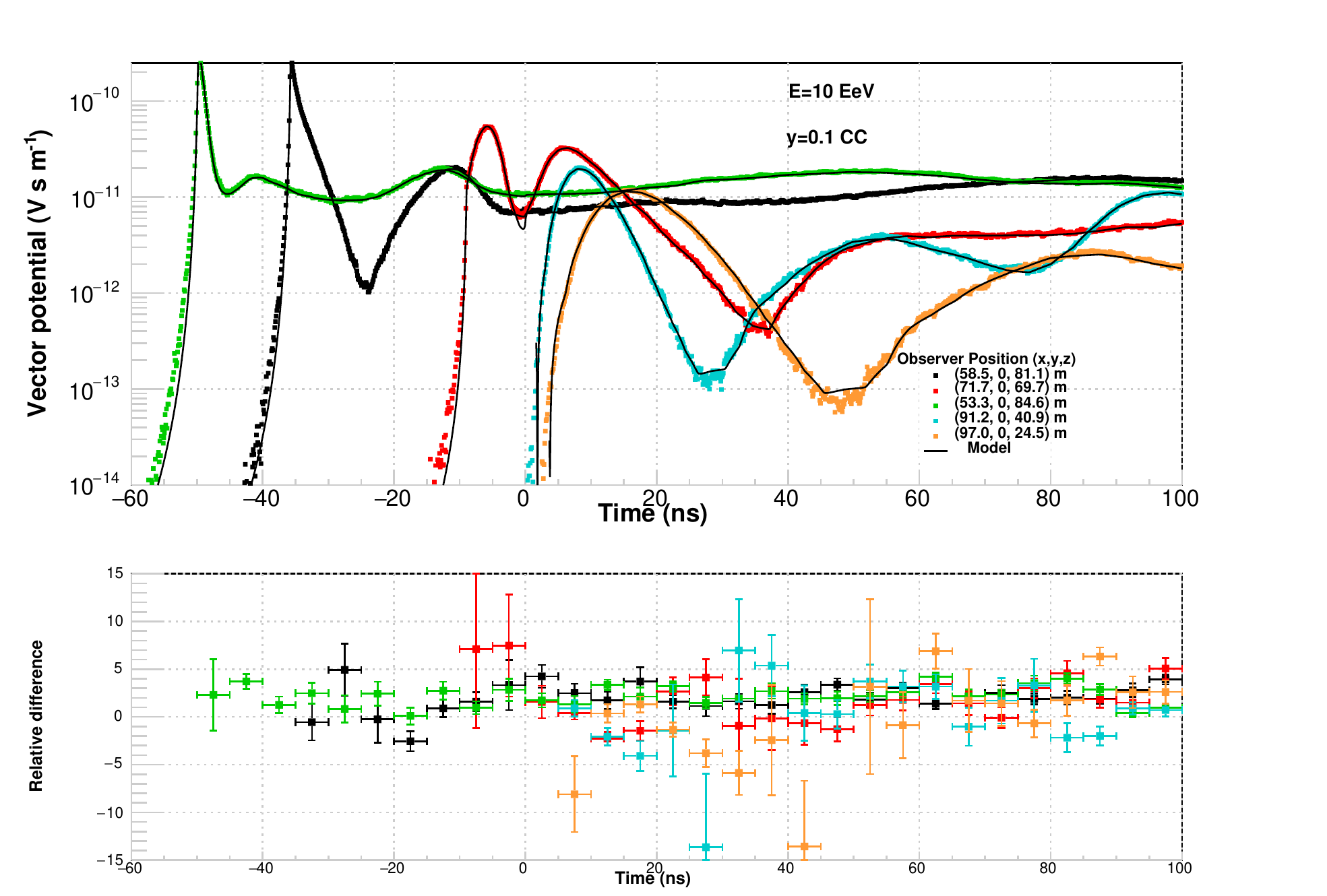}}
&\resizebox*{0.51\textwidth}{!}{\includegraphics{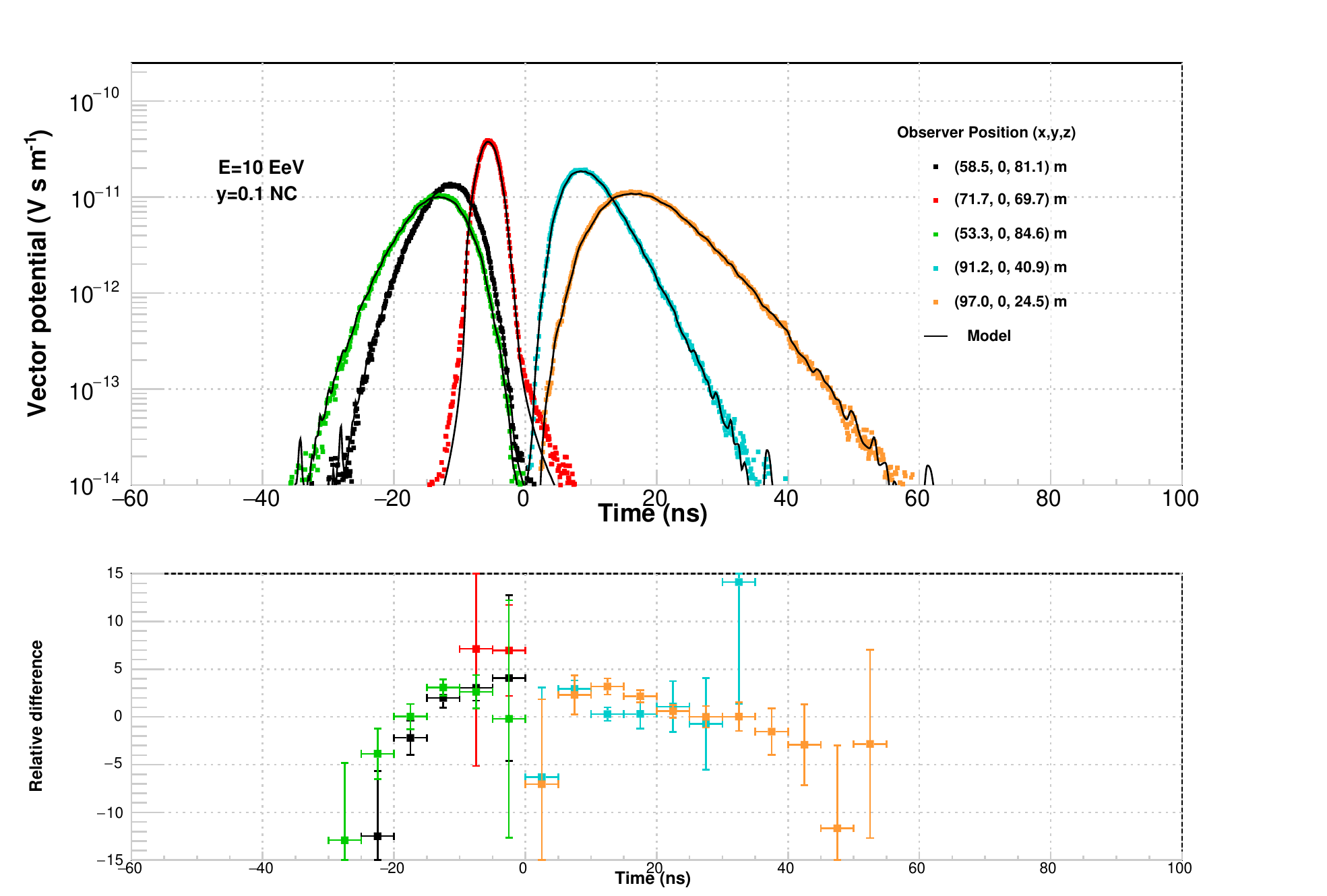}}
\\
\end{tabular}\par}
\caption{
Comparison of the vector potential obtained in this approach with full simulations for four different cases: 
Charged-current $\nu_e$ interactions on the left and neutral-current interactions (of any flavor) on the right, 
for 0.1 EeV neutrino energy at the top and 10 EeV at the bottom. In all cases the energy fraction transferred to the nucleus 
is $y=0.1$. In each case we display two panels: The upper panel gives the amplitude of the vector potential for five
observers located at different positions, all 100 m away from the starting point of the shower
(see diagram in Fig.~\ref{fig:observers}). The solid black lines are the values obtained with the approach discussed 
in this article (Section \ref{SS:algorithm}) while the colored dots correspond to the values directly obtained in full ZHAireS simulations. 
The lower panel in each case gives the relative difference between the full ZHAireS simulations and the results
obtained with the approach of this article, normalized to the simulation results and averaged in bins
of 5 ns width. The error bars reproduce the rms fluctuations within the bin.
The accuracy of the model worsens at times when the amplitude of the vector potential also drops significantly with respect to the value at the peak and is not expected to be relevant for practical purposes.
}
\label{fig:vp_nue_CC_nu_NC_y0p1}
\end{figure*}

In Fig.\,\ref{fig:vp_nue_CC_nu_NC_y0p1} we show the vector potential for showers induced in $\nu_e$ CC (left) and $\nu$ NC (right) interactions, for two neutrino energies ($E_\nu=0.1$ EeV in the top panels, 10 EeV in the bottom ones) fixing the momentum transferred to the nucleus to the value $y=0.1$. For the NC the vector potentials are very similar, basically scaled by a factor of order 100, that is by the energy ratio. For the CC case we note that the vector potential has larger amplitude because essentially all the neutrino energy is converted into the mixed type shower. For the 10 EeV CC neutrino electron interaction the development of the electron shower, that carries $90\%$ of the neutrino energy, displays a characteristic multi-peak structure because of the LPM effect. A similar multi-peak structure is apparent in the vector potential as anticipated. The differences between this approach and full ZHAireS simulations are of order $\pm 5\%$ in the region near the peak, dropping to $\sim 20\%$ at the onset and at the latest times of the pulse (for the same reasons explained above) where the amplitude is negligible. 

\begin{figure*}[h]
{\centering 
\begin{tabular}{cc}
\resizebox*{0.51\textwidth}{!}{\includegraphics{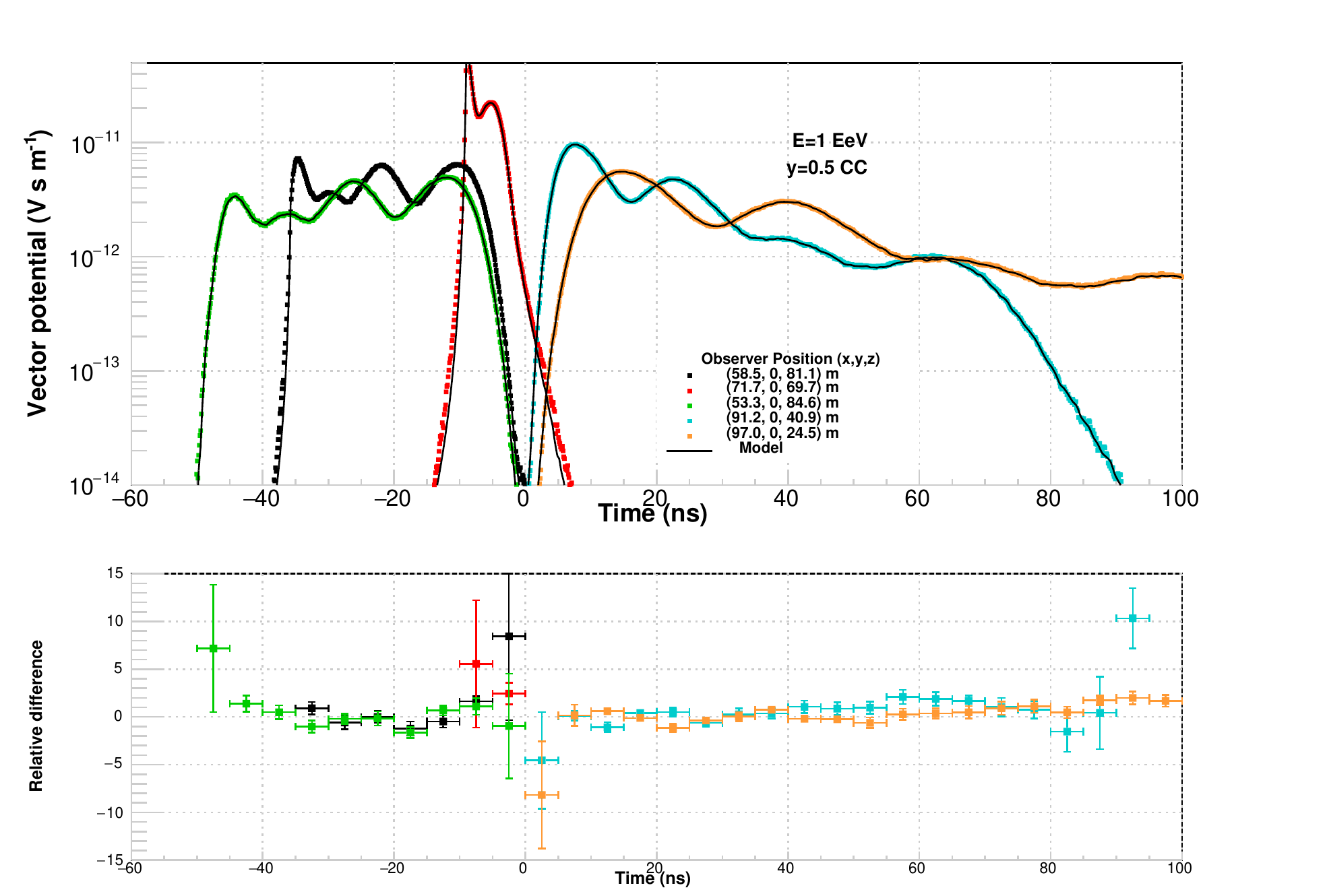}}
&\resizebox*{0.51\textwidth}{!}{\includegraphics{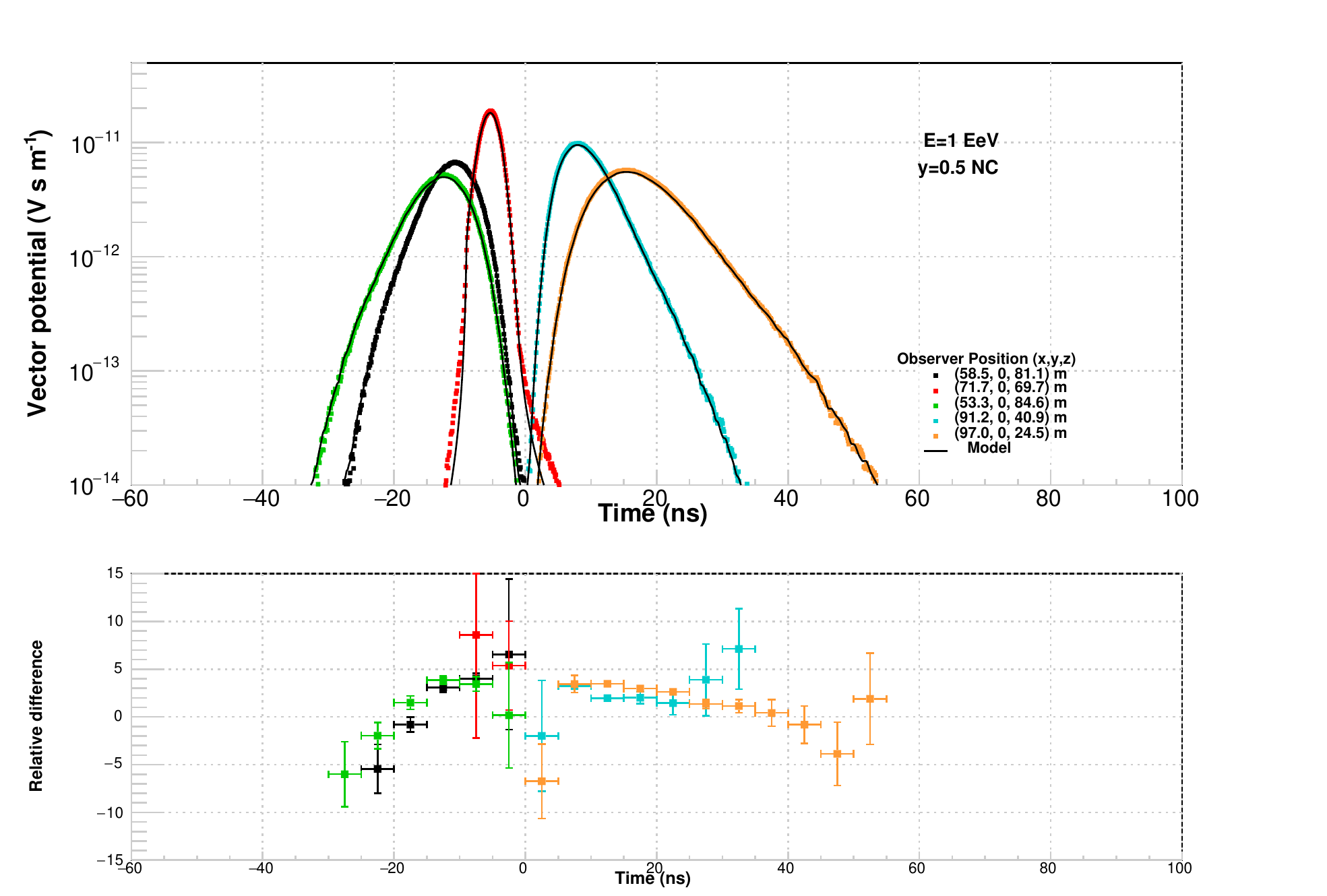}}
\\
\resizebox*{0.51\textwidth}{!}{\includegraphics{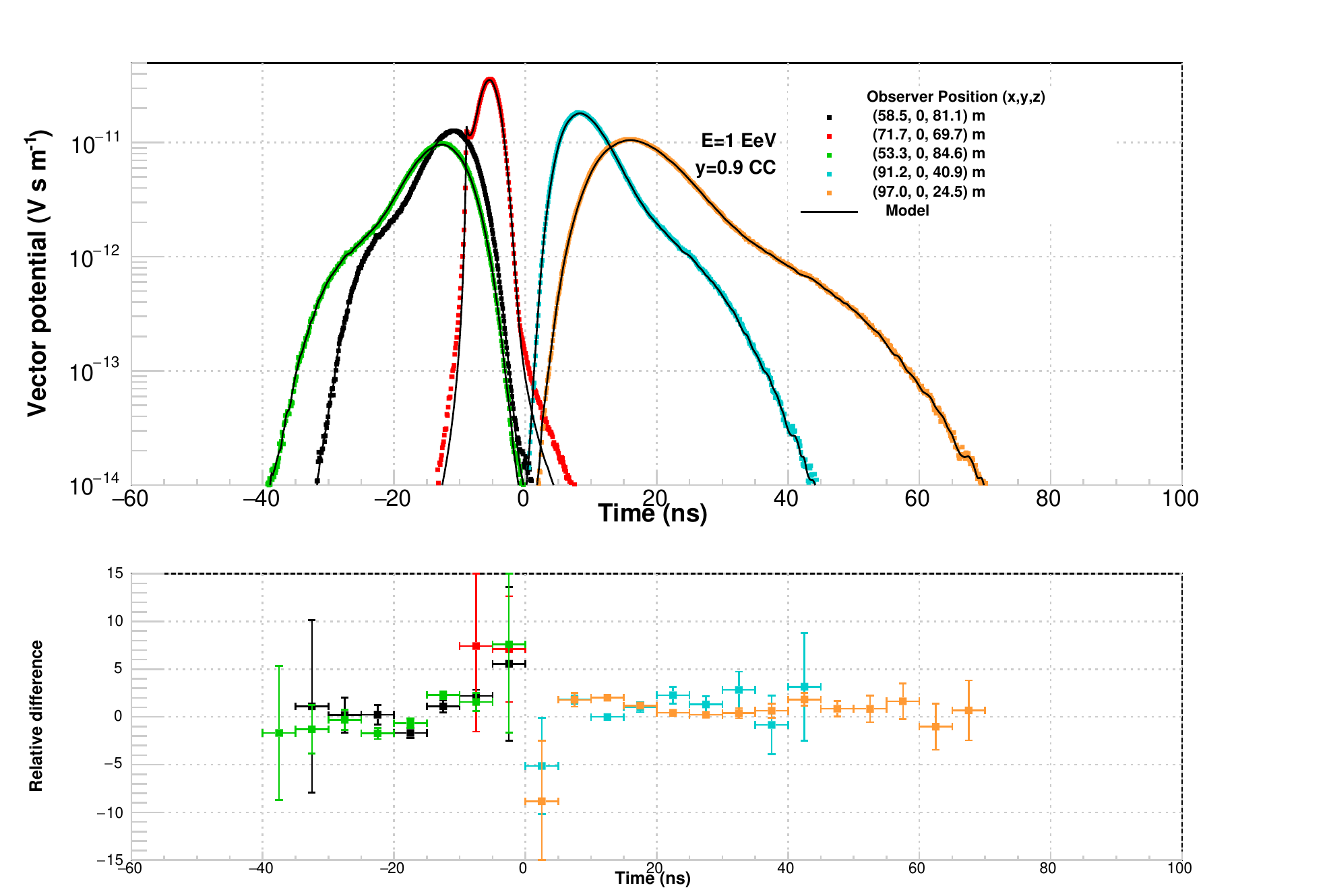}}
&\resizebox*{0.51\textwidth}{!}{\includegraphics{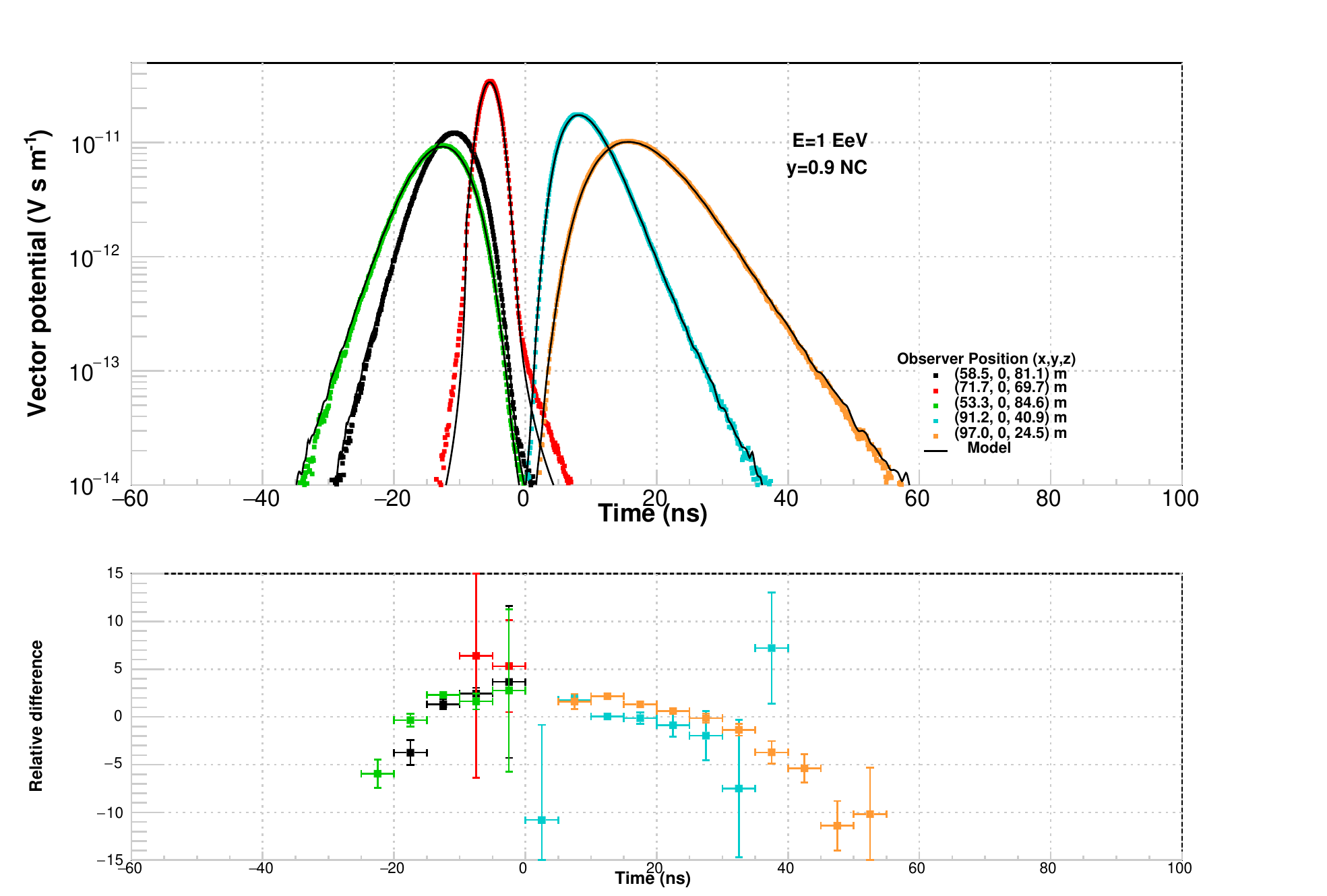}}
\\
\end{tabular}\par}
\caption{Comparison of the vector potential obtained with the approach of this article and with full simulations for showers due to $\nu_e$ charged-current interactions (left panels) and by all flavour $\nu$ neutral-current interactions (right panels). All panels are for a fixed neutrino energy of 1 EeV and two different values of the energy transfer, $y=0,5$ and $y=0.9$ as labelled. The observers are located at the same positions and the color coding is the same as in Fig.~\ref{fig:vp_nue_CC_nu_NC_y0p1}. 
For each vector potential we also show the relative difference between the full ZHAireS simulations and the approach in this article, normalized to the ZHAireS results and averaged in bins of 5 ns width. The rms value of the average difference in each bin is also shown. The accuracy of the model worsens at times when the amplitude of the vector potential also drops significantly with respect to the value at the peak and is not expected to be relevant for practical purposes.
}
\label{fig:vp_nue_CC_NC_yvar}
\end{figure*}

In Fig.\,\ref{fig:vp_nue_CC_NC_yvar} we repeat the comparison of the vector potential obtained using our methodology to those obtained with full simulations for both $\nu_e$ CC (top panels) and $\nu$ NC (bottom panels) interactions, this time for a fixed neutrino energy  $E_\nu=$1 EeV and two different values of  the fractional energy transfer, $y=0.5$ and $y=0.9$. 
The features of the vector potential at the chosen observer positions are qualitatively very similar to those described in Fig.~\ref{fig:vp_nue_CC_nu_NC_y0p1}. Naturally, as the value of $y$ increases, less energy is transferred to the electron shower and the LPM effect becomes less apparent. For NC interactions the amplitude of the vector potential continues to scale with shower energy which in this case is proportional to $y$. Again, the differences between our methodology and full ZHAireS simulations for the same showers is $\sim 3\%$ in the region near the peak, dropping to $\sim 20\%$ at the earliest and latest phases where the amplitude is negligible. 

\begin{figure*}[h]
{\centering 
\begin{tabular}{cc}
\resizebox*{0.51\textwidth}{!}{\includegraphics{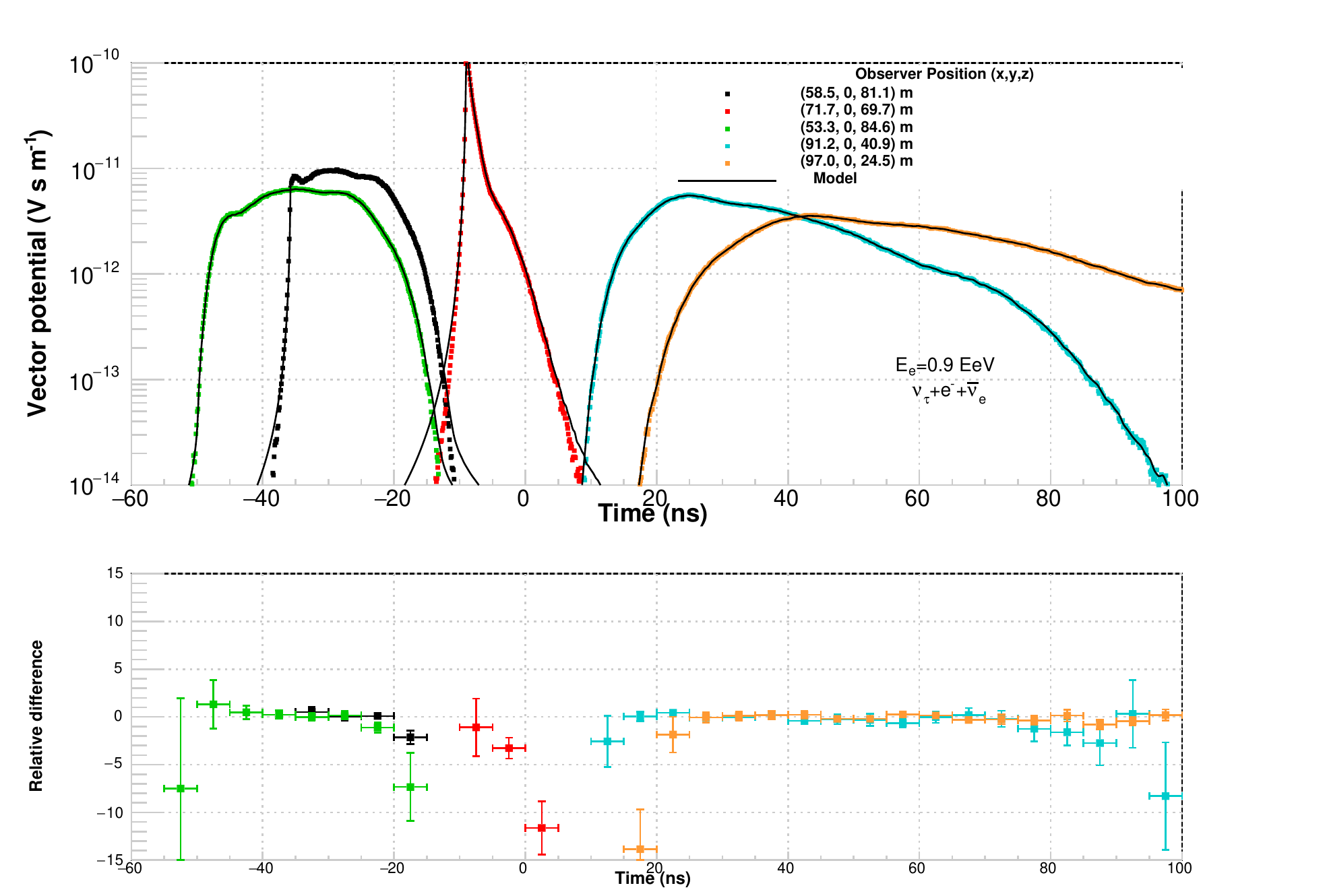}}
&\resizebox*{0.51\textwidth}{!}{\includegraphics{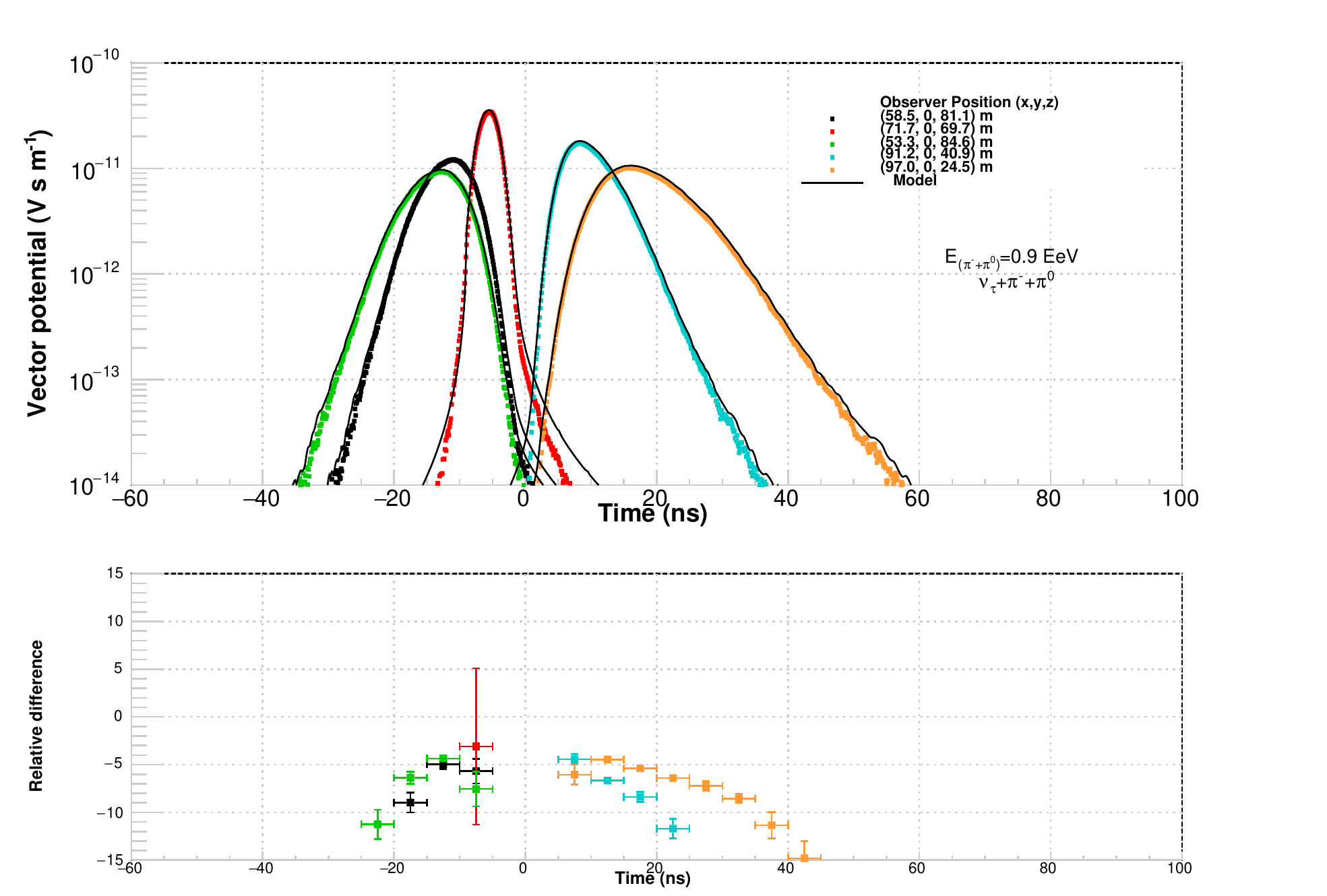}}
\\
\end{tabular}\par}
\caption{The vector potential in showers in homogeneous ice induced by the decay products of a $\tau$ lepton produced in a $\nu_\tau$ CC interaction of energy $E_\nu=$1 EeV. In the right panel we consider the decay channel $\tau^-\rightarrow e^-+\bar\nu_e+\nu_\tau$ with the 
electron carrying an energy $E_e=0.9$ EeV, while in 
the left panel the decay channel is $\tau^-\rightarrow \pi^- + \pi^0 + \nu_\tau$ with the sum of the energy of the pions being $E=0.9$ EeV. The solid black lines represent the vector potentials as obtained with our model, applying the algorithm explained in Section \ref{SS:algorithm}, while the colored lines correspond to the vector potential obtained in full ZHAireS simulations of the same showers. The vector potential was obtained for observers at several positions with respect to the starting point of the shower, with all the observers at a distance of $\sim\,$100 m to that point, although at different angles with respect to the shower axis.
For each vector potential we also show the relative difference between the full ZHAireS simulations and the approach in this article, normalized to the ZHAireS results and averaged in bins of 5 ns width. The rms value of the average difference in each bin is also shown. The accuracy of the model worsens at times when the amplitude of the vector potential also drops significantly with respect to the value at the peak and is not expected to be relevant for practical purposes.
}
\label{fig:vp_tau_decay}
\end{figure*}

In Fig.\,\ref{fig:vp_tau_decay} we again compare the vector potentials obtained in this approach with those obtained in full ZHAireS simulations for showers induced by the decay of a $\tau$ lepton. The tau lepton is produced by $\nu_\tau$ CC interactions for neutrinos of energy $E_\nu=$1 EeV. In the left panel we consider the case of a $\tau$ lepton decaying into an electron of energy $E_e=0.9$ EeV (and a $\bar\nu_e$ along with a $\nu_\tau$ that are assumed not to interact), while in the right panel the $\tau$ decays hadronically to a charged and a neutral pion (as well as a $\nu_\tau$ that does not interact), where the sum of the energies of the pions is $0.9$ EeV. Similarly to the other cases, the agreement between this model and the simulations of the same showers is quite good, at the $3\%$ level except at the earliest and latest time of the pulses. Naturally, the decay into an electron leads to a vector potential with multi-peak structure.  

Finally, in Fig.\,\ref{fig:vp_nu_NC_distance} we explore the accuracy of this semi-analytical calculation as the observer distance changes for a $\nu$ NC interaction at energy $E_\nu=1$ EeV and $y=0.1$. The observers are located at distances measured from the start of the shower of 4.13~m, 41.3~m, 413~m and 4.13~km at an angle of $76^\circ$ to the shower axis. The accuracy of the approach becomes again $3\%$ for a distance greater than $\sim 4$~m. Again, at the onset of the pulse there is some decrease of accuracy but only in a region in which the vector potential is so small that it can be neglected. 

The $\pm 5\%$ accuracy achieved with our semi-analytical approach, that assumes that the shape of lateral 
charge distribution $f(r',z')$ depends weakly on $z'$ and neglects the component of the form factor
along the line of sight to the observer, justifies the use of these approximations.

%

\section{Conclusions}
\label{S:Conclusions}

We have extended the semi-analytical method described in  \cite{ARZ11} to neutrino interactions of different types. We have been able to reproduce the pulses generated by neutral and charged-current interactions of all neutrino flavors in homogeneous ice with a simple approach that effectively accounts for the lateral distribution of the shower with a form factor that depends on a single variable. Using this form factor it is possible to describe the vector amplitude and the radio-pulse for any practical geometry. 
To obtain numerically the form factor we rely on shower simulations and calculate the vector potential in the Fraunhofer limit for observers located in the direction of the Cherenkov angle. These simulations have been performed using HERWIG for the neutrino interaction code, TAUOLA for the simulation of the tau decay and ZHAireS for the simulation of the shower and the vector potential of the associated radio pulse. 

We have extended the calculations of the form factor for electromagnetic showers~\cite{ARZ11} to showers initiated by hadron fragments such as those produced in the fragmentation of the nucleus or from the decays of $\tau$. We have parameterized the electromagnetic and hadronic form factors, updating the parameterization of the electromagnetic form factor given in \cite{ARZ11}, and exploring their accuracy. 
We have classified the showers produced by different flavour neutrinos and for different interactions as linear combinations of these two shower types. We have compared the form factors obtained with ZHAireS for the neutrino interactions to those obtained by linear combinations of the  electromagnetic and hadronic form factors, showing that the obtained parameterizations agree with the full simulations within a few percent in a region of $\pm 0.5$ ns around the peak where the amplitude is highest.

Finally, we have determined that the vector potential obtained in this approach, using the form factors and the longitudinal development of the showers $Q(z')$ as inputs, reproduces that obtained from the full ZHAireS simulations. This exercise has been repeated for different neutrino energies, neutrino flavours, interaction types, different fractional energy transfers to the nucleus and different distances from the start of the shower. We have studied the accuracy of the approach for the main part of the pulse in all these cases which has been shown to be better than $\sim \pm 5\%$ even for distances as close as 4~m from the start of the shower. We have also noted that the accuracy becomes worse at the onset and end of the pulses, reaching $\sim 20\%$ when the amplitude of the vector potential drops by at least three orders of magnitude relative to its maximum value. This is due to the smaller accuracy of the fits to the electromagnetic and hadronic vector potentials in Eqs.\,(\ref{eq:vp_fit_EM}) and (\ref{eq:vp_fit_HAD}) in the time range outside the window $(-0.5,0.5)$ ns. An improved fit or an interpolation of the MC results instead of the fit is expected to increase the accuracy of our methodology even in the region where the amplitude of the vector potential is negligible compared to that at the peak. 

Applications of this approach are envisaged for the design and study of different experimental arrangements designed to trigger on pulses induced by high energy neutrinos and thus to study their performance. The technique is also expected to be of interest for the analysis of data from such experimental arrangements. 
The semi-analytical approach to describe any type of neutrino  shower just requires two functions of a single variable, the form factors of electromagnetic and hadronic showers, to accurately describe the electric field amplitude at practically any relevant position of an antenna relative to the shower. The radio pulse can be evaluated numerically using these functions once the neutrino flavor, interaction type and fraction of energy transferred to the nucleus are specified. 

In spite of using the universal parameterizations of the form factors, the profiles of the showers have to be known. These can be obtained through full simulations but the calculation only needs to be done once. Calculating the pulse at an arbitrary position is then extremely easy and fast. Storing shower profiles of the electromagnetic and hadronic showers is relatively easy to do. A library of showers can be used to make lengthy simulations that accurately sample the performance of experimental facilities in an efficient way. 
This makes our methodology convenient for the implementation in more detailed studies of detector layout optimization. An example is the recently developed NuRadioMC code \cite{NuRadioMC_2019}, in which the approach explained in this work has been implemented, including a library of simulated shower profiles in ice. The goal is to develop cost-effective detectors for UHE neutrino fluxes, which will in turn address one of the most fundamental unanswered questions in astroparticle physics. 

\section{Acknowledgments}
We thank M. Seco for computing assistance. 
P.M.H thanks Subsidios para Viajes y/o Estad\'ias, Facultad de Ingenier\'ia, Universidad Nacional de La Plata, 
Argentina. P.M.H. further thanks the Astroparticle Physics group at the Univ. of Santiago de Compostela 
for hospitality.
J.A-M. and E.Z. thank the financial support of Ministerio de Econom\'\i a, Industria y Competitividad 
(FPA 2017-85114-P), Xunta de Galicia (ED431C 2017/07) and RENATA Red Nacional Tem\'atica de Astropart\'\i culas 
(FPA2015-68783-REDT).
This work is supported by the Mar\'\i a de Maeztu Units of Excellence program MDM-2016-0692 
and the Spanish Research State Agency.
This work is co-funded by the European Regional Development Fund (ERDF/FEDER program).



\begin{thebibliography}{99}

\bibitem{NS_Merger2017}
Abbott, B. P., {\sl et al.}
Gravitational Waves and Gamma-rays from a Binary Neutron
Star Merger: GW170817 and GRB 170817A
Astrophys. J. Lett. {\bf 848} 2 (2017) L13.

\bibitem{NSFollowUp}
Abbott, B. P., {\sl et al.}
Multi-messenger Observations of a Binary Neutron Star
Merger
Astrophys. J. Lett. {\bf 848} 2 (2017) L12.

\bibitem{NS_NuSearch}
Albert, A., {\sl et al.}
Search for High-energy Neutrinos from Binary Neutron
Star Merger GW170817 with ANTARES, IceCube, and the Pierre Auger Observatory
Astrophys. J. Lett. {\bf 850} 2 (2017) L35.



\bibitem{Nagano-Watson_2000}
M. Nagano, and A.~A. Watson, 
Observations and implications of the ultrahigh-energy cosmic rays,
Rev. Mod. Phys. {\bf 72} (2000) 689.

\bibitem{Dawson_2017}
B.~R. Dawson, M. Fukushima, and P. Sokolsky,
Past, present, and future of UHECR observations,
Prog. Theor. Exp. Phys. {\bf 12} (2017) A101. 

\bibitem{Mollerach-Roulet_2018}
S. Mollerach, and E. Roulet,
Progress in high-energy cosmic ray physics,
Prog. in Part. and Nucl. Phys. {\bf 98} (2018) 85.

\bibitem{IceCube_PRL2014}
M.~G. Aartsen {\it et al.} [IceCube Collab.],
Observation of High-Energy Astrophysical Neutrinos in Three Years of IceCube Data,
Phys. Rev. Lett. {\bf 113} (2014) 101101.

\bibitem{IceCube_ApJ_2016}
M.~G. Aartsen {\it et al.} [IceCube Collab.],
Observation and characterization of a cosmic muon neutrino flux from the Northern Hemisphere using six years of IceCube data,
Astrophys. J. {\bf 833} (2016) 3. 

\bibitem{Becker_PhysRep_2008}
J.~K. Becker, 
High-energy neutrinos in the context of multimessenger physics
Phys. Rep. {\bf 458}, 173 (2008).

\bibitem{Anchordoqui_review_2014}
L.~A. Anchordoqui et al.,
Cosmic neutrino pevatrons: A brand new pathway to astronomy, astrophysics, and particle physics,
Journal of High Energy Astrophysics, {\bf 1}, 1 (2014).

\bibitem{Veronique_review_2011}
B. Baret, and V. Van Elewyck,
High-energy neutrino astronomy: detection methods and first achievements,
Rep. Prog. Phys. {\bf 74} (2011) 046902.

\bibitem{Alvarez-Muniz_ICRC17}
J. Alvarez-Mu\~niz, 
Ultra-high energy neutrinos: status and prospects,
PoS (ICRC2017) 1111 and refs. therein.

\bibitem{AERA}
B. Pont for the Pierre Auger Collaboration,
A large radio detector at the Pierre Auger Observatory - measuring the properties of cosmic rays up to the highest energies, 
in Procs. of the $36^{\rm th}$ International Cosmic Ray Conference 2019,
PoS(ICRC2019)395.

\bibitem{GRAND} 
J. Alvarez-Mu\~niz {\it et al.} [GRAND Collaboration],
The Giant Radio Array for Neutrino Detection (GRAND): Science and Design, 
Sci. China-Phys. Mech. Astron. {\bf 63}, 219501 (2020).

\bibitem{ARIANNA_2014}
S.~W. Barwick {\it et al.} [ARIANNA Collab.], 
Design and Performance of the ARIANNA HRA-3 Neutrino Detector Systems,
IEEE Trans. Nucl. Sci. {\bf 62}, 2202 (2015);
S.~W. Barwick {\it et al.} [ARIANNA Collab.],
A first search for cosmogenic neutrinos with the ARIANNA Hexagonal Radio Array, 
Astropart. Phys. {\bf 70}, 12 (2015).

\bibitem{ARA_PRD2016}
P. Allison {\it et al.} [ARA Collab.], 
Performance of two Askaryan Radio Array stations and first results in the search for ultrahigh energy neutrinos, 
Phys. Rev. D {\bf 93}, 082003 (2016). 

\bibitem{Connolly-Vieregg_Review_2017}
A.~L. Connolly and A.~G. Vieregg,
Radio Detection of High Energy Neutrinos,
Neutrino Astronomy: current status, future prospects, pp. 217-240 (2017)
World Scientific. arXiv:1607.08232v1 [astro-ph.HE].

\bibitem{Askaryan62} 
G.~A. Askar'yan, 
Excess Negative Charge of an Electron-Photon Shower and its Coherent Radio Emission,
{\sl Soviet Physics} JETP {\bf 14}, 441 (1962);
G.~A. Askar'yan,
Coherent Radio Emission from Cosmic Showers in Air and in Dense Media,
JETP {\bf 48}, 988 (1965).

\bibitem{Auger_nus_JCAP2019}
A. Aab {\it et al.} [Pierre Auger Collab.],
Probing the origin of ultra-high-energy cosmic rays with neutrinos in the EeV energy range using the Pierre Auger Observatory,
JCAP {\bf 10}, 022 (2019).

\bibitem{Auger_point_JCAP2019}
A. Aab {\it et al.} [Pierre Auger Collab.],
Limits on point-like sources of ultra-high-energy neutrinos with the Pierre Auger Observatory,
JCAP {\bf 11}, 004 (2019).

\bibitem{IceCube_PRD2018}
M.~G. Aartsen {\it et al.} [IceCube Collab.],
Differential limit on the EHE cosmic neutrino flux in the presence of astrophysical background from nine years of IceCube data,
Phys. Rev. D {\bf 98} (2018) 062003. 

\bibitem{ANITA_2019}
P.~W. Gorham {\it et al.} [ANITA Collaboration],
Constraints on the ultra-high energy cosmic neutrino flux from the fourth flight of ANITA,
Phys. Rev. D {\bf 99}, 122001 (2019).

\bibitem{IceCube_TXS}
M.~G. Aartsen {\it et al.} [IceCube Collab.], 
Multimessenger observations of a flaring blazar coincident with high-energy neutrino IceCube-170922A, 
Science {\bf 361} (2018) 146. 


\bibitem{ZHS92} 
E. Zas, F. Halzen, T. Stanev, 
Electromagnetic pulses from high-energy showers: Implications for neutrino detection,
Phys. Rev. D {\bf 45}, 362 (1992).

\bibitem{Saltzberg_SLAC_sand} 
D.~Saltzberg {\it et al.}, 
Observation of the Askaryan Effect: Coherent Microwave Cherenkov Emission from Charge Asymmetry in High-Energy Particle Cascades,
Phys.\ Rev.\ Lett.\ {\bf 86}, 2802 (2001).

\bibitem{Gorham_SLAC_salt} 
P.~W. Gorham {\it et al.}, 
Accelerator measurements of the Askaryan effect in rock salt: A roadmap toward teraton
underground neutrino detectors,
Phys. Rev. D {\bf 72}, 023002 (2005). 

\bibitem{Gorham_SLAC_ice} 
P.~W. Gorham {\it et al.}, 
Observations of the Askaryan Effect in Ice,
Phys. Rev. Lett. {\bf 99}, 171101 (2007). 

\bibitem{Miocinovic_SLAC_sand}
P.~Miocinovic {\it et al.}, 
Time-domain measurement of broadband coherent Cherenkov radiation,
Phys. Rev. D {\bf 74}, 043002 (2006).  

\bibitem{ARZ10}
J.~Alvarez-Mu\~ niz, A. Romero-Wolf, E. Zas, 
Cherenkov radio pulses from electromagnetic showers in the time domain,
Phys. Rev. D {\bf 81}, 123009 (2010).

\bibitem{ARZ11}
J. Alvarez-Mu\~ niz, A. Romero-Wolf, E. Zas, 
Practical and accurate calculations of Askaryan radiation
Phys. Rev. D {\bf 84}, 103003 (2011).

\bibitem{RICE03}
I. Kravchenko {\it et al.}, 
RICE limits on the diffuse ultrahigh energy neutrino flux,
Phys.\ Rev.\ D\ {\bf 73}, 082002 (2006).

\bibitem{ANITA_long_2009}
P.~W. Gorham {\it et al.} [ANITA Collab.],
The Antarctic Impulsive Transient Antenna ultra-high energy neutrino detector: Design, performance, and sensitivity for the 2006–2007 balloon flight, Astropart. Phys. {\bf 32}, 10 (2009).

\bibitem{LUNASKA1}
T.~H. Hankins, R.~D. Ekers, J.~D. O'Sullivan, A search ̆for lunar radio Cherenkov emission from high-energy neutrinos, Mon. Not. R. Astron. Soc. {\bf 283}, 1027 (1996).

\bibitem{GLUE}
P.~W. Gorham, {\it et al.}, Experimental limit on the cosmic diffuse ultrahigh energy neutrino flux, Phys. Rev. Lett. {\bf 93}, 041101 (2004).

\bibitem{Zhelez_moon}
A.~R. Beresnyak, R.~D. Dagkesamanskii, I.~M. Zheleznykh, A.~V. Kovalenko, V.~V. Oreshko, Limits on the flux of ultrahigh-energy neutrinos from radio astronomical observations, Astron. Rep. {\bf 49}, 127 (2005).

\bibitem{LUNASKA2}
C.W. James, {\it et al.}, LUNASKA experiments using the Australia Telescope Compact Array to search for ultrahigh energy neutrinos and develop technol- ogy for the lunar Cherenkov technique, Phys. Rev. D {\bf 81}, 042003 (2010).

\bibitem{RESUN}
] T.R. Jaeger, R.L. Mutel, K.G. Gayley, Project RESUN, a radio EVLA search for UHE neutrinos, Astropart. Phys. {\bf 34}, 293 (2010).

\bibitem{Westerbork}
S. Buitink, {\it et al.}, Constraints on the flux of ultra-high energy neutrinos from Westerbork Synthesis Radio Telescope observations, Astron. $\&$ Astrophys. {\bf 521}, 1 (2010).

\bibitem{Parkes}
J.~D. Bray, {\it et al.}, Limit on the ultrahigh-energy neutrino flux from lunar observations with the Parkes radio telescope, Phys. Rev. D {\bf 91}, 063002 (2015).

\bibitem{Bray_APP_2016}
J.~D. Bray,
The sensitivity of past and near-future lunar radio experiments to ultra-high-energy cosmic rays and neutrinos, Astropart. Phys. {\bf 77}, 1 (2016), and refs. therein.

\bibitem{ZHAireS_ice} 
J. Alvarez-Mu\~niz, W.~R. Carvalho Jr., M. Tueros, and E. Zas,
Coherent Cherenkov radio pulses from hadronic showers up to EeV energies,
Astropart. Phys. {\bf 35}, 287 (2012).

\bibitem{LPM}
L.~Landau, I.~Pomeranchuk, 
Electron-Cascade Processes at Ultra-High Energies,
{\sl Dokl.\ Akad.\ Nauk\ SSSR} {\bf 92}, 735 (1935);
A.B.~Migdal, 
Bremsstrahlung and Pair Production in Condensed Media at High Energies,
Phys.\ Rev.\ {\bf 103}, 1811 (1956). 

\bibitem{Stanev_LPM}
T.~Stanev {\it et al.},
Development of ultrahigh-energy electromagnetic cascades in water and lead including the Landau-Pomeranchuk-Migdal effect,
Phys. Rev. D {\bf 25}, 1291 (1982).

\bibitem{Buniy_PRD2001}
R.~V.~Buniy, J.~P.~Ralston,
Radio detection of high energy particles: Coherence versus multiple scales,
Phys. Rev. D {\bf 65}, 016003 (2001).

\bibitem{Hanson_APP2017}
J.~C. Hanson, A.~L. Connolly,
Complex analysis of Askaryan radiation: A fully analytic treatment including the LPM effect and Cascade Form Factor,
Astropart. Phys. {\bf 91}, 75 (2017).

\bibitem{alz97} 
J. Alvarez-Mu\~niz, E. Zas, 
Cherenkov radio pulses from EeV neutrino interactions: the LPM effect,
Phys. Lett. B {\bf 411}, 218 (1997).

\bibitem{alz98} 
J. Alvarez-Mu\~niz, E. Zas, 
The LPM effect for EeV hadronic showers in ice: implications for radio detection of neutrinos,
Phys. Lett. B {\bf 434}, 396 (1998).

\bibitem{alvz99}
J. Alvarez-Mu\~niz, R.A. V\'azquez and E. Zas,
Characterization of neutrino signals with radiopulses in dense media through the Landau-Pomeranchuk-Migdal effect,
Phys. Rev. D {\bf 61}, 023001 (1999).

\bibitem{Chen_2012}
C.-Y. Hu, C.-C. Chen, P. Chen,
Near-field effects of Cherenkov radiation induced by ultra high energy cosmic neutrinos,
Astropart. Phys. {\bf 35}, 421 (2012).

\bibitem{ZHS_calculations_PRD2013}
D. Garc\'\i a-Fern\'andez, J. Alvarez-Mu\~niz, W.~R. Carvalho, A. Romero-Wolf, E.Zas, 
Calculations of electric fields for radio detection of ultrahigh energy particles,
Phys. Rev. D {\bf 87}, 023003 (2013).

\bibitem{Jackson}
J.~D. Jackson, 
{\it Classical Electrodynamics 3rd Ed.},
Wiley, New York, (1998).

\bibitem{Jackson_AmJPhys}
J.~D. Jackson, 
From Lorenz to Coulomb and other explicit gauge transformations,
Am. J. Phys. {\bf 70}, 917 (2002).

\bibitem{Connolly_nu_xsection_PRD2011}
A.~L. Connolly, R.~S. Thorne and D. Waters,
Calculation of high energy neutrino-nucleon cross sections and uncertainties using the Martin-Stirling-Thorne-Watt parton distribution functions and implications
for future experiments,
Phys. Rev. D {\bf 83}, 113009 (2011).

\bibitem{HERWIG}
G. Corcella, I.~G. Knowles, G. Marchesini, S. Moretti, K.
Odagiri, P. Richardson, M.~H. Seymour, and B. R. Webber,
HERWIG 6: an event generator for hadron emission reactions with 
interfering gluons (including supersymmetric processes),
J. High Energy Phys. {\bf 01}, 010 (2001).

\bibitem{TAUOLA}
S. Jadach, Z. Was, R. Decker, and J.~H. K\"uhn, 
The $\tau$ decay library TAUOLA, version 2.4,
Comput. Phys. Commun. {\bf 76}, 361 (1993),

\bibitem{AIRESManual}
 S.~J. Sciutto, 
 AIRES User's Manual and Reference Guide; version 2.6.0
(2002), available electronically at {\tt www.fisica.unlp.edu.ar/auger/aires}.

\bibitem{TIERRAS}
M. Tueros, and S. Sciutto, 
TIERRAS: a package to simulate high energy cosmic ray showers underground,
underwater and under-ice,
Comp. Phys. Comm. {\bf 181}, 30 (2010).

\bibitem{ZHAireS_air} 
J. Alvarez-Mu\~niz, W.~R. Carvalho Jr., and E. Zas,
Monte Carlo simulations of radio pulses in atmospheric showers using ZHAireS,
Astropart. Phys. {\bf 35}, 325 (2012).

\bibitem{Saftoiu_salt_2019}
A. Saftoiu,
Estimation of radio emission from neutrino induced showers in rock salt above $10^{18}$ eV,
Astropart. Phys. {\bf 113}, 22 (2019).

\bibitem{InvisibleEnergy}
A. Aab {\sl et al.} (Pierre Auger Collaboration), 
Data-driven estimation of the invisible energy of cosmic ray showers with the Pierre Auger Observatory,
Phys. Rev. D {\bf 100} (2019) no.8, 082003.

\bibitem{Feynman_1963}
R.~P. Feynman, R.~B. Leighton, M. Sands, 
{\it The Feynman Lectures in Physics}. Addison-Wesley, (1963).

\bibitem{NuRadioMC_2019}
C. Glaser {\it et al.},
NuRadioMC: Simulating the radio emission of neutrinos from interaction to detector,
Eur. Phys. J. C {\bf 80}, 77 (2020).

\end{thebibliography}
\end{document}